\PassOptionsToPackage{table,xcdraw}{xcolor}
\documentclass[11pt,a4paper]{article}
\pdfoutput=1
\usepackage{jheppub}
\usepackage{gensymb}
\DeclareSymbolFont{matha}{OML}{txmi}{m}{it}
\DeclareMathSymbol{\varv}{\mathord}{matha}{118}
\usepackage{amsmath}
\usepackage{amssymb}
\usepackage{graphicx}
\usepackage{subfigure}
\usepackage{pstricks}
\usepackage{bm}
\usepackage{pbox}
\usepackage{placeins}
\usepackage{booktabs}
\usepackage[T1]{fontenc}
\usepackage{footnote}
\usepackage{pdfpages}
\usepackage{hhline}
\usepackage{multirow}
\usepackage{multicol}
\usepackage{enumitem}
\usepackage[toc,page]{appendix}
\allowdisplaybreaks
\usepackage{comment}
\usepackage{float}
\usepackage{slashed}
\usepackage{xcolor}
\usepackage{textcomp}
\usepackage{gensymb}
\usepackage{multirow}
\usepackage[numbers]{natbib}
\usepackage{notoccite}
\usepackage{tabu}
\usepackage[normalem]{ulem}

\usepackage{rotating}
\usepackage{tabularx}
\usepackage[labelfont=bf]{caption} 

\FloatBarrier

\usepackage[many]{tcolorbox}

\renewcommand{\thetable}{\Roman{table}}

\definecolor{MyDarkBlue}{rgb}{0.1, 0.1, 0.8} 
\definecolor{MyLightBlue}{rgb}{0.22,0.51,0.9}
\definecolor{MyGreen}{rgb}{0.0, 0.5, 0.0}
\definecolor{BrickRed}{rgb}{0.8, 0.25, 0.33}

\RequirePackage{hyperref}
\hypersetup{colorlinks, citecolor=MyGreen,linkcolor=cyan, urlcolor=MyLightBlue}

\makeatletter
\gdef\@fpheader{}
\makeatother
\begin{document}
\title{\bf Viable quark-lepton Yukawa ratios and nucleon decay predictions in $SU(5)$ GUTs with type-II seesaw}

\author[]{Stefan Antusch,}
\author[]{Kevin Hinze,}
\author[]{and Shaikh Saad}

\affiliation[]{Department of Physics, University of Basel, Klingelbergstrasse\ 82, CH-4056 Basel, Switzerland}

\emailAdd{stefan.antusch@unibas.ch, kevin.hinze@unibas.ch, shaikh.saad@unibas.ch}
\abstract{
We investigate the viability of predictive schemes for quark-lepton Yukawa ratios and nucleon decay in non-supersymmetric SU(5) Grand Unified Theories (GUTs) where neutrino masses are generated by a type II seesaw mechanism. The scalar sector of the considered scenario contains 5-, 24- and 45-dimensional representations plus a 15-dimensional representation for realising the type II seesaw. Predictions for the ratios of the quark and lepton Yukawa couplings emerge when the relevant entries of the Yukawa matrices are generated from single joint GUT operators (i.e.\ under the condition of \textit{single operator dominance}).  Focusing on the 2nd and 3rd family and hierarchical Yukawa matrices, we show that only two sets of predictions, $\frac{y_\tau}{y_b}=\frac{3}{2}$, $\frac{y_\mu}{y_s}=\frac{9}{2}$ and $\frac{y_\tau}{y_b}=2$, $\frac{y_\mu}{y_s}=6$ are viable. To further investigate both options, we extend the minimal scenarios to two ``toy models'', including the 1st family of charged fermions,
and calculate the models' predictions, e.g.\ for the nucleon decay rates and for the masses of the light relics that are potentially within reach of colliders.
}
\maketitle

\section{Introduction}
Grand Unified Theories (GUTs)~\cite{Pati:1974yy, Georgi:1974sy, Georgi:1974yf, Georgi:1974my, Fritzsch:1974nn} present a fascinating framework for physics beyond the Standard Model (SM). On top of the unification of gauge couplings, they (partially) unify the fermions into joint GUT representations, providing an interesting framework to address the flavor puzzle, i.e. the origin of the observed fermion masses, mixing angles and CP violating phases. 

The simplest model based on the unifying group $SU(5)$, proposed by Georgi and Glashow in 1974\ \cite{Georgi:1974sy}, predicts between the charged lepton and down-type quark Yukawa matrices the GUT scale relation 
\begin{align}\label{eq-Ye=YdT}
Y_e=Y_d^T,
\end{align}
which in particular results in a GUT scale unification of tau and bottom Yukawa couplings, as well as a unification of muon and strange Yukawa couplings. It is however well-known that these results disagree with the low energy experimental data of fermion masses, as long as only the known SM particles are taken into account in the renormalization group (RG) evolution\  \cite{Arason:1991ic}. 
Typical approaches for dealing with this shortcoming can be subdivided into two classes: 
\begin{enumerate}
\item In the first class of approaches any GUT prediction for the GUT scale quark-lepton Yukawa coupling ratios is removed by introducing new degrees of freedom, which basically make these GUT scale ratios freely adjustable parameters. This can be achieved, for example, by adding additional renormalizable (see e.g.\  \cite{Dorsner:2014wva, Tsuyuki:2014xja, FileviezPerez:2016sal, Saad:2019vjo, Dorsner:2019vgf, Dorsner:2021qwg}) or non-renormalizable (see e.g.\ \cite{Ellis:1979fg, Dorsner:2006hw, Bajc:2006ia, Bajc:2007zf}) GUT operators, such that the Yukawa couplings emerge as linear combinations from these operators. While, on the one hand, consistency with the low energy observable data can be restored this way, on the other hand the predictivity for the Yukawa ratios is lost.
\item The second class of approaches maintains predictivity for the GUT scale Yukawa ratios by considering models which feature alternative GUT relations. They have in common that only a single GUT operator dominates each relevant entry in the GUT Yukawa matrices, a concept we refer to as \textit{single operator dominance}\ (cf.\ \cite{Antusch:2009gu, Antusch:2013rxa,Antusch:2019avd}). Different types of GUT operators can result in different GUT predictions for the respective quark-lepton Yukawa ratio. Their consistency with the low energy experimental data imposes constraints on the model parameters, such as the value of the GUT scale or the couplings to additional fields introduced e.g.\ for explaining the observed light neutrino masses.  
\end{enumerate} 

We follow in this paper an approach of the second type. With each Yukawa entry being dominated by a different single non-renormalizable GUT operator, different GUT scale charged lepton and down-type quark ratios (cf.\ \cite{Antusch:2009gu, Antusch:2013rxa,Antusch:2019avd}) can be considered in the different Yukawa entries, maintaining a high predictivity while potentially having a better viability with the experimental data.\footnote{Examples of models employing the concept of \textit{single operator dominance} can be found, e.g. in\  \cite{Antusch:2012fb, Antusch:2013kna, Antusch:2017ano, Antusch:2018gnu}.} It has been recently suggested~\cite{Antusch:2021yqe}, that in non-SUSY $SU(5)$ GUTs the GUT scale ratios 
\begin{align}\label{eq-ytau/yb=3/2}
\frac{y_\tau}{y_b}=\frac{3}{2}\qquad\text{and}\qquad \frac{y_\mu}{y_s}=\frac{9}{2},
\end{align}
where $y_i$ denotes the low-scale singular value of the Yukawa matrices, and $i$ its fermion flavor, are in good agreement with the experimental data, if the neutrino masses originate from a type I seesaw mechanism\ \cite{Minkowski:1977sc}. We will in this paper show that on top of the GUT scale relations in Eq.~\eqref{eq-ytau/yb=3/2}, the ratios 
\begin{align}\label{eq-ytau/yb=2}
\frac{y_\tau}{y_b}=2\qquad\text{and}\qquad \frac{y_\mu}{y_s}=6
\end{align}
can also be viable if instead the neutrino masses are generated by a type II seesaw mechanism\ \cite{Magg:1980ut,Schechter:1980gr} in a non-SUSY $SU(5)$ GUT. Both of these possibilities of GUT scale charged lepton and down-type quark Yukawa ratios can be realized if the Georgi-Glashow model is extended such that the Higgs sector contains the representations \textbf{5}, \textbf{24} and \textbf{45}. Moreover, adding a 15-dimensional Higgs representation allows for a type II seesaw mechanism\ \cite{Dorsner:2005fq}.

In this paper we investigate the viability of the two combinations of GUT scale Yukawa coupling ratios listed in Eq.~\eqref{eq-ytau/yb=3/2} and Eq.~\eqref{eq-ytau/yb=2} in a GUT scenario in which the neutrino masses are generated by a type II seesaw mechanism and the Higgs sector consists of the representations \textbf{5}, \textbf{24}, \textbf{45} and \textbf{15}.\footnote{A model with the same Higgs sector was already considered in\ \cite{Dorsner:2007fy}. However, the focus of the present work lies on the viability of the predicted GUT scale Yukawa ratios.} Extending this GUT scenario to two ``toy models'' (following \cite{Antusch:2014poa}) we compute the nucleon decay rates for 13 different decay channels using the \texttt{ProtonDecay} package\ \cite{Antusch:2020ztu} and calculate the allowed ranges of the masses of the added intermediate-scale scalar fields, hinting various possibilities to test our models by future experiments\  \cite{Acciarri:2015uup, Abe:2018uyc, An:2015jdp}. 

The paper is organised as follows: In Section~\ref{sec-model} we introduce our GUT scenario and investigate the viability of the GUT scale ratios $\frac{y_\tau}{y_b}=\frac{3}{2}$ and $\frac{y_\mu}{y_s}=\frac{9}{2}$, respectively the GUT scale relations $\frac{y_\tau}{y_b}=2$ and $\frac{y_\mu}{y_s}=6$, w.r.t. the experimental low-scale fermion masses. We further extend our GUT scenario to two ``toy models'' both predicting one of these relations. We explain the numerical procedure in Section~\ref{sec-numerical analysis}, before discussing the numerical results in Section~\ref{sec-results} and concluding in Section~\ref{sec-conclusions}. We list the renormalization group equations (RGEs) of the gauge and Yukawa couplings in Appendix~\ref{app-rge gauge and yukawa} and show in Appendix~\ref{app-RGE kappa} how the added scalar fields change the running of the effective neutrino mass operator. Finally, in Appendix~\ref{sec-perturbativity of Ydelta above the GUT scale} we discuss the perturbativity of our ``toy models'' by approximating the running of the neutrino couplings above the GUT scale.

\section{Model}\label{sec-model}
\subsection{Particle content}\label{sec-particle content}
We investigate an extension of the Georgi-Glashow (GG) model\ \cite{Georgi:1974sy} in which neutrino masses are generated by a type II seesaw mechanism\ \cite{Magg:1980ut} and in which charged lepton and down-type quark Yukawa ratios are predicted at the GUT scale through the concept of \emph{single operator dominance} \cite{Antusch:2009gu, Antusch:2013rxa}. On top of the usual GG particle content, this SU(5) GUT scenario contains the Higgs field $\bm{15_H}$ to allow for a type II seesaw mechanism\ \cite{Dorsner:2005fq}. Furthermore, a Higgs field $\bm{45_H}$ is used in order to allow for gauge coupling unification and to produce some of the operators which predict fixed GUT scale charged lepton and down-type quark Yukawa ratios\ \cite{Antusch:2009gu, Antusch:2013rxa}. In summary, the Higgs sector consists of the representations $\bm{5_H}$, $\bm{24_H}$, $\bm{45_H}$, and $\bm{15_H}$, whereas the fermion sector is given by ${\bm{\overline{5}_F}}_i$, and ${\bm{10_F}}_i$, where $i=1,2,3$. We will show that within this scenario, depending on which particles are light and which are heavy, two different combinations of GUT scale Yukawa ratios are possible: (i) $\frac{y_\tau}{y_b}=\frac{3}{2}$ and $\frac{y_\mu}{y_s}=\frac{9}{2}$, respectively (ii) $\frac{y_\tau}{y_b}=2$ and $\frac{y_\mu}{y_s}=6$. 

As usual, the GUT representations ${\bm{\overline{5}_F}}_i$, and ${\bm{10_F}}_i$ contain the SM fermions
\begin{align}
&{\bm{\overline{5}_F}}_i= \ell_i (1,2,-\frac{1}{2}) \oplus d_i^c (\overline{3},1,\frac{1}{3}),  \label{GG1}\\
&{\bm{10_F}}_i= q_i (3,2,\frac{1}{6}) \oplus u_i^c (\overline{3},1,-2/3) \oplus e_i^c (1,1,1). \label{GG2} 
\end{align} 
Under the SM gauge group $SU(3)_C\times SU(2)_L\times U(1)_Y$ the Higgs fields have the following decomposition:
\begin{align}
{\bm{5_H}} =&\; T_1(\overline{3},1,-\frac{1}{3}) \oplus H_1(1,2,\frac{1}{2}),\\
{\bm{45_H}} =&\; \phi_1(8,2,\frac{1}{2}) \oplus \phi_2(6,1,-\frac{1}{3}) \oplus \phi_3(3,3,-\frac{1}{3}) \oplus \phi_4(\overline{3},2,-\frac{7}{6}) \nonumber\\
&\oplus \phi_5(\overline{3},1,\frac{4}{3}) \oplus T_2(3,1,-\frac{1}{3}) \oplus H_2(1,2,\frac{1}{2}),  \label{GG4}\\
{\bm{24_H}}=&\; \Sigma_8 (8,1,0) \oplus \Sigma_1 (1,3,0) \oplus \Sigma_0 (1,1,0) \oplus \Sigma_3(3,2,-\frac{5}{6}) \oplus \Sigma_{\overline{3}}(\overline{3},2,\frac{5}{6}), \label{GG5}\\
\bm{15_H}=&\;\Delta_1\left(1,3,1\right)\oplus\Delta_3(3,2,\frac{1}{6})\oplus\Delta_6(\overline{6},1,-\frac{2}{3}).
\end{align} 
A linear combination of $H_1$ and $H_2$ gives the electroweak breaking Higgs doublet $h$. In the following we will assume that the second Higgs doublet $h^\perp$ which is given by a linear combination of $H_1$ and $H_2$ orthogonal to $h$ has its mass at the GUT scale $M_\text{GUT}$. Moreover, the color triplets $T_1$ and $T_2$ mix and we denote the corresponding mass eigenstates by $t_1$ and $t_2$. Due to nucleon decay restrictions the masses of these two fields have to be above $10^{11}$\ GeV \cite{Nath:2006ut}.

\subsection{Yukawa sector}\label{sec-Yukawa sector}
\subsubsection{Possible Yukawa couplings}\label{sec-possible Yukawa couplings}
With the particle content specified above the following Yukawa couplings are possible
\begin{align}
    (Y_{\bm{\overline{5}}})_{ij}: \quad {\bm{10_F}}_i{\bm{\overline{5}_F}}_j\,\bm{X}\quad &\supset \quad (Y_d)_{ij},\,(Y_e)_{ij},\label{eq:5F10F}\\
    (Y_{\bm{10}})_{ij}: \quad {\bm{10_F}}_i{\bm{10_F}}_j\,\bm{Y}\quad &\supset \quad (Y_u)_{ij},\\
    \frac{1}{\sqrt{2}}(Y_{\Delta})_{ij}: \quad {\bm{\overline{5}_F}}_i{\bm{\overline{5}_F}}_j\,\bm{Z}\quad &\supset \quad  \frac{1}{\sqrt{2}}(Y_1)_{ij},\,(Y_3)_{ij},\, \frac{1}{\sqrt{2}}(Y_6)_{ij},\label{eq:5F5F}
\end{align}
where $\bm{X}$, $\bm{Y}$, and $\bm{Z}$ represent one or multiple Higgs fields and where $Y_u$, $Y_d$, and $Y_e$ are the typical SM up-type, down-type and charged lepton Yukawa couplings, whereas $Y_1$, $Y_3$, and $Y_6$ are new quasi-like Yukawa couplings defined by the interactions $\frac{1}{\sqrt{2}}(Y_1)_{ij} \Delta_1 \ell_i\ell_j$, $(Y_3)_{ij} \Delta_3 d^c_i\ell_j$, and $\frac{1}{\sqrt{2}}(Y_6)_{ij} \Delta_6 d^c_i d^c_j$. These quasi-Yukawa matrices are allowed to have entries of the order of $\mathcal{O}(1)$. Moreover, the SU(5) operator ${\bm{10_F}}_i{\bm{\overline{5}_F}}_j\,\bm{X}$ generates after the GUT symmetry breaking interactions between the SM fermions and the Higgs fields $\phi_1$, $\phi_2$, $\phi_3$, $\phi_4$, and $\phi_5$. For all of these interactions, the largest entry in the corresponding quasi-Yukawa matrix is of the order of $\mathcal{O}(10^{-2})$, since the quasi-Yukawa matrices are up to Clebsch-Gordan (GC) coefficients identical to $Y_d$. Furthermore, if $\bm{Y}$ contains $\bm{5_H}$, the operator ${\bm{10_F}}_i{\bm{10_F}}_j\,\bm{Y}$ only generates the up-type Yukawa coupling. If $\bm{Y}$ was further allowed to also contain $\bm{45_H}$, then couplings between the SM fermions and the Higgs fields $\phi_1$, $\phi_2$, $\phi_3$, $\phi_4$, and $\phi_5$ of the order of $\mathcal{O}(1)$ would be allowed. We will in the following assume that there exists a symmetry which forbids the latter case. 

\subsubsection{Neutrino mass generation}\label{sec-neutrino mass generation}
After the GUT symmetry breaking the following interactions are relevant for the neutrino mass generation\ \cite{Magg:1980ut}:
\begin{align}
    \mathcal{L}\supset - \frac{1}{\sqrt{2}}Y_1 \ell \ell \Delta_1 - \mu hh\Delta_1 - M_{\Delta_1} \Delta_1^\dagger \Delta_1,
\end{align}
where $Y_1$ stems from the GUT operator defined in Eq.~\eqref{eq:5F5F}, $\mu$ is a trilinear scalar coupling   coming from the GUT operator $\bm{5_H}\bm{5_H}\bm{15_H}$, and $M_{\Delta_1}$ describes the mass of the scalar field $\Delta_1$. Integrating out $\Delta_1$ below its mass $M_{\Delta_1}$ gives the effective dimension five neutrino mass operator $\kappa$
\begin{align}
    \kappa=-2\frac{\mu Y_1 }{M_{\Delta_1}^2},
\end{align}
from which the neutrino mass matrix $m_\nu$ is obtained by 
\begin{align}
    m_\nu=-\frac{v^2}{4}\kappa,
\end{align}
after the electroweak breaking Higgs doublet $h$ takes its vacuum expectation value $v=246\,$GeV\ \cite{ParticleDataGroup:2020ssz}.

\subsubsection{Viable GUT scale Yukawa ratios}\label{sec-viable GUT scale Yukawa ratios}
From now on we assume that (i) the Yukawa matrices $Y_{\bm{\overline{5}}}$ and $Y_{\bm{10}}$ are hierarchical and (ii) the concept of \emph{single operator dominance}  in the Yukawa sector. The latter assumption means that each Yukawa entry is dominated by a singlet GUT operator. With these two assumptions the charged lepton and down-type quark masses stem dominantly from the GUT operators $\mathcal{O}_2$, and $\mathcal{O}_3$ for the second, and third family, respectively, if the GUT scale Yukawa matrix $Y_{\bm{\overline{5}}}$ is given by
\begin{align}
Y_{\bm{\overline{5}}}=
\begin{pmatrix}
0 & 0 & 0 \\
0 & \mathcal{O}_2 & 0 \\
0 & 0 & \mathcal{O}_3
\end{pmatrix}
+\dots \;,
\end{align}
where the dots represent subdominant contributions. On the one hand, if the operators $\mathcal{O}_2$, and $\mathcal{O}_3$ are chosen as (specifying the contraction between different brackets by the SU(5) representation in the index position)
\begin{align}
&\mathcal{O}_2 = \left({\bm{10_F}}_2 \bm{45_H}^*\right)_{\bm{5}}\left({\bm{5_F}}_2 \bm{24_H}\right)_{\bm{\overline{5}}},    \\
&\mathcal{O}_3 = \left({\bm{10_F}}_3 \bm{5_H}^*\right)_{\bm{45}}\left({\bm{5_F}}_3 \bm{24_H}\right)_{\bm{\overline{45}}},    
\end{align}
then the GUT scale Yukawa ratios $\frac{y_\tau}{y_b}=\frac{3}{2}$ and $\frac{y_\mu}{y_s}=\frac{9}{2}$ are predicted. On the other hand, a second interesting choice of the operators $\mathcal{O}_2$ and $\mathcal{O}_3$ giving rise to the GUT scale Yukawa ratios $\frac{y_\tau}{y_b}=2$ and  $\frac{y_\mu}{y_s}=6$ is given by
\begin{align}
&\mathcal{O}_2 = \left({\bm{10_F}}_2 \bm{45_H}^*\right)_{\bm{5}}\left({\bm{5_F}}_2 \bm{24_H}\right)_{\bm{\overline{5}}},    \\
&\mathcal{O}_3 = \left({\bm{10_F}}_3 \bm{45_H}^*\right)_{\bm{5}}\left(\bm{24_H}\right)_{\bm{\overline{5}}\otimes\bm{5}}\left({\bm{5_F}}_3 \bm{1_H}\right)_{\bm{\overline{5}}}.    
\end{align}
In the following we will show that the two possibilities for GUT scale charged lepton and down-type quark Yukawa ratios presented above are indeed viable if the particle content introduced in Section~\ref{sec-particle content} is assumed. For this analysis we investigate the RGE evolution of the SM parameters together with the quasi-Yukawa couplings $Y_1$, $Y_3$ and $Y_6$. At 1-loop the RGEs of the charged lepton and down-type quark Yukawa couplings read (cf. Appendix~\ref{app-rge gauge and yukawa})  
\begin{align}
&16\pi^2 \mu \frac{dY_d}{d\mu}=\beta_d^{SM} + \left(\mathcal{H}(\mu,m_{\Delta_3})Y_3Y^\dagger_3\right)Y_d + \Big(\mathcal{H}(\mu,m_{\Delta_6})2Y_6Y^\ast_6 \Big)Y_d, 
\\
&16\pi^2 \mu \frac{dY_e}{d\mu}=\beta^{SM}_e + Y_e \left( \frac{3}{2}\mathcal{H}(\mu,m_{\Delta_1})Y^\ast_1Y_1  \right) + Y_e \left( \frac{3}{2}\mathcal{H}(\mu,m_{\Delta_3})Y^\dagger_3Y_3  \right),
\end{align}
where $\beta_d^{SM}$ and $\beta_e^{SM}$ are the SM beta functions, and where $\mathcal{H}(\mu,m)$ is the Heaviside step function defined by
\begin{align}\label{eq-Heaviside step function}
    \mathcal{H}(\mu,m)=\left\{
\begin{array}{lll}
1, && \mu>m \\
0, && \, \mu\leq m \\
\end{array}.
\right. 
\end{align} 
Now, considering a fixed GUT scale and comparing the running in the SM with the running in which also the effects of $Y_1$, $Y_3$ and $Y_6$ are taken into account, we make the following observations for the GUT scale Yukawa charged lepton and down-type quark Yukawa ratios: If $m_{\Delta_6}$ is well below the GUT scale, while $m_{\Delta_1}$ is around the GUT scale, smaller GUT scale charged lepton and down-type quark Yukawa ratios become viable due to the term $\left(2Y_6Y^\ast_6 \right)Y_d$ ``speeding up'' the running of $Y_d$. Similarly, if $m_{\Delta_1}$ is well below the GUT scale, while $m_{\Delta_6}$ is around the GUT scale, larger GUT scale charged lepton and down-type quark Yukawa ratios become viable due to the term $Y_e \left( \frac{3}{2}Y^\ast_1Y_1  \right)$ ``speeding up'' the running of $Y_e$. 

\begin{figure}
    \centering
    \includegraphics[width=10cm]{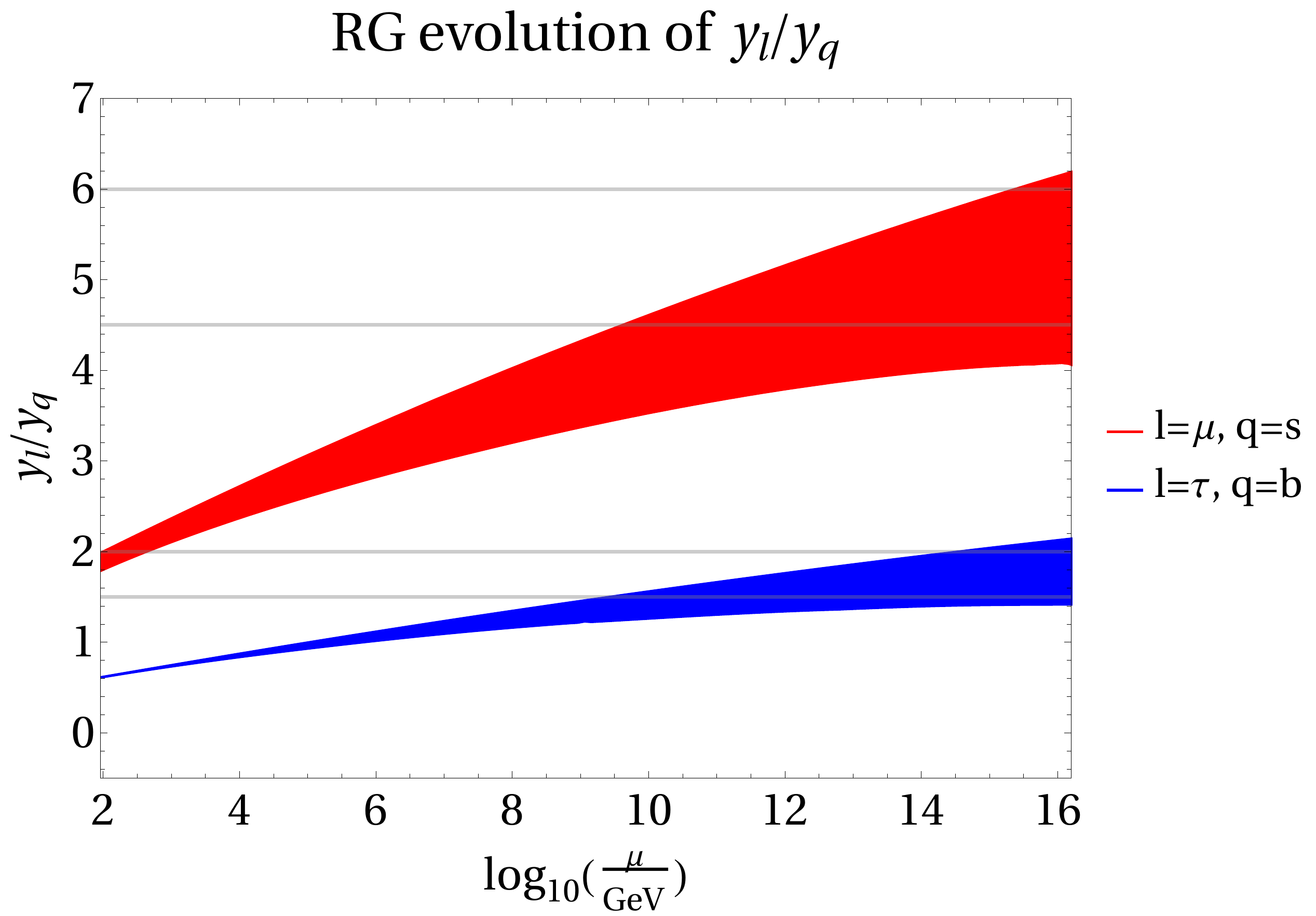}
    \caption{Estimated RG evolution of the 1-$\sigma$ ranges of the Yukawa ratios $y_\tau/y_b$ (blue) and $y_\mu/y_s$ (red) taking into account the effect of $Y_1$, $Y_3$ and $Y_6$. The horizontal grid lines indicate the values $\frac{3}{2}$, 2, $\frac{9}{2}$ and 6.}\label{fig-yukawa ratio running}
\end{figure}

In order to investigate the viability of the GUT scale Yukawa ratios $\frac{y_\tau}{y_b}=\frac{3}{2}$ and $\frac{y_\mu}{y_s}=\frac{9}{2}$, respectively $\frac{y_\tau}{y_b}=2$ and $\frac{y_\mu}{y_s}=6$, we considered the bottom-up RG evolution performing a Markov chain Monte Carlo (MCMC) analysis for which we varied the SM parameters around their experimental central value and weighted the points according to the $\chi^2$-function, while also varying the masses $m_{\Delta_1}$, $m_{\Delta_3}$ and $m_{\Delta_6}$. As argued in Appendix~\ref{sec-perturbativity of Ydelta above the GUT scale}, we stopped the integration if the largest singular value of $Y_1$, $Y_3$ and $Y_6$ got larger than 1.9. We then computed on a narrow grid the 1-$\sigma$ HPD intervals of the two ratios $\frac{y_\tau}{y_b}$ and $\frac{y_\mu}{y_s}$. Figure~\ref{fig-yukawa ratio running} shows that for GUT-scales above $10^{15.5}$\,GeV within the 1-$\sigma$ uncertainties one of the two possibilities of GUT-scale Yukawa ratios $\frac{y_\tau}{y_b}=\frac{3}{2}$ and $\frac{y_\mu}{y_s}=\frac{9}{2}$, or $\frac{y_\tau}{y_b}=2$ and $\frac{y_\mu}{y_s}=6$ can be realized.

\subsection{Toy models}\label{sec-toy models}
After having argued that there are two viable possibilities for GUT scale charged lepton and down-type quark Yukawa ratios for the second and third generation, namely (i) $\frac{y_\tau}{y_b}=\frac{3}{2}$ and $\frac{y_\mu}{y_s}=\frac{9}{2}$, and (ii) $\frac{y_\tau}{y_b}=2$ and $\frac{y_\mu}{y_s}=6$, we now extend the two scenarios to ``toy models'' which also include the first family. Assuming the concept of \emph{single operator dominance}, Table~\ref{tab-dominant operators} shows which operator is dominant in which entry of the Yukawa matrices $ Y_{\bm{\overline{5}}}$, $ Y_{\bm{10}}$, and $ Y_{\Delta}$ which were defined in Section~\ref{sec-possible Yukawa couplings}.

\begin{table}[h!]
    \centering
    \renewcommand{\arraystretch}{1.2}
    \begin{tabu}{ccc}
    \tabucline[1.1pt]{-}
       operator  & model 1 & model 2 \\
       \hline
       $(Y_\Delta)_{ij}$ & ${\bm{\overline{5}_F}}_i{\bm{\overline{5}_F}}_j\bm{15_H}$ & ${\bm{\overline{5}_F}}_i{\bm{\overline{5}_F}}_j\bm{15_H}$ \\
       $(Y_{\bm{10}})_{ij}$ & ${\bm{10_F}}_i{\bm{10_F}}_j{\bm{5_H}}$ & ${\bm{10_F}}_i{\bm{10_F}}_j{\bm{5_H}}$ \\
       $(Y_{\bm{\overline{5}}})_{11}$ & $({\boldsymbol{10_F}}_1\boldsymbol{5_H}^*)_{\boldsymbol{5}}(\boldsymbol{24_H})_{\boldsymbol{24}}(\boldsymbol{1_H})(\boldsymbol{24_H})_{\boldsymbol{24}}({\boldsymbol{\overline{5}_F}}_1\boldsymbol{1_H})_{\bm{\overline{5}}}$ & $({\boldsymbol{10_F}}_{1} \boldsymbol{45_H}^*)_{\boldsymbol{\overline{45}}}({\boldsymbol{\overline{5}_F}}_1 \boldsymbol{24_H})_{\boldsymbol{45}}$ \\
       $(Y_{\bm{\overline{5}}})_{22}$  & $({\bm{10_F}}_2{\bm{45_H}^*})_{\bm{5}}({\bm{\overline{5}_F}}_2\bm{24_H})_{\bm{\overline{5}}}$ & $({\bm{10_F}}_2{\bm{24_H}})_{\bm{10}}({\bm{\overline{5}_F}}_2\bm{5_H})_{\bm{\overline{10}}}$  \\
       $(Y_{\bm{\overline{5}}})_{33}$  & $({\bm{10_F}}_3{\bm{5_H}^*})_{\bm{45}}({\bm{\overline{5}_F}}_3\bm{24_H})_{\bm{\overline{45}}}$ & $({\bm{10_F}}_3{\bm{45_H}^*})_{\bm{5}} (\bm{24_H})_{\bm{24}} ({\bm{\overline{5}_F}}_3\bm{1_H})_{\bm{\overline{5}}}$  \\
       \tabucline[1.1pt]{-}
    \end{tabu}
    \caption{Dominant operators in the respective Yukawa entries in the two toy models.}
    \label{tab-dominant operators}
\end{table}
From the renormalizable operator ${\bm{\overline{5}_F}}{\bm{\overline{5}_F}}\bm{15_H}$ we get the GUT scale relation 
\begin{align}
    Y_1=Y_3=Y_6.
\end{align}
Moreover, since $Y_{\bm{10}}$ is dominated by the operator ${\bm{10_F}}{\bm{10_F}}{\bm{5_H}}$, the up-type Yukawa matrix is symmetric, i.e. $Y_u=Y_u^T$. We assume that the operator ${\bm{10_F}}{\bm{10_F}}{\bm{45_H}}$, which would yield anti-symmetric contributions, vanishes as a consequence of an appropriate family symmetry (see e.g.\ \cite{Antusch:2021yqe} for an example of such a symmetry).
Furthermore, at the GUT scale  the Yukawa matrices $Y_e$ and $Y_d$ are in the two ``toy models'' related via\ \cite{Antusch:2009gu, Antusch:2013rxa}
\begin{align}
&\text{model 1:}\qquad Y_e=\text{diag}\left(
\frac{4}{9},\frac{9}{2},\frac{3}{2}\right)\cdot Y_d^T, \label{eq-GUT scale Yukawa ratio 1}\\   
&\text{model 2:}\qquad Y_e=\text{diag}\left(
\frac{1}{2},6,2\right)\cdot Y_d^T.\label{eq-GUT scale Yukawa ratio 2}
\end{align}

\section{Numerical analysis}\label{sec-numerical analysis}
\subsection{Implementation}\label{sec-implementation}
The models 1 and 2 defined in Section~\ref{sec-toy models} are implemented at the GUT scale. Since they only differ in their corresponding charged lepton sector, all other sectors can be implemented identically in both of the two ``toy models''. The symmetric Yukawa matrices $Y_u$ and $Y_\Delta$ can be diagonalized by a Takagi decomposition: 
\begin{align}
    Y_u &= U_{u}^\dagger Y_u^\text{diag} U_{u}^*, \\
    Y_\Delta &= U_{\Delta} Y_\Delta^\text{diag} U_{\Delta}^T,
\end{align}
where $U_u$ and $U_\Delta$ are unitary matrices and $Y_u^{\text{diag}}$ as well as $Y_\Delta^{\text{diag}}$ are defined as
\begin{align}
    Y_u^\text{diag} &= \text{diag}(y_1^u,y_2^u,y_3^u), \\
    Y_\Delta^\text{diag} &= \text{diag}(y_1^\Delta,y_2^\Delta,y_3^\Delta),
\end{align}
with $y_i^u$ and $y_i^\Delta$ being real entries larger or equal to zero for $i=1,2,3.$ The unitary matrices $U_g$ ($g=u,\Delta$) can be parametrized by the physical parameters $\theta_{12}^g$, $\theta_{13}^g$, $\theta_{23}^g$, $\delta^g$, $\beta_1^g$, and $\beta_2^g$ as well as the unphysical parameters $\alpha_1^g$, $\alpha_2^g$, and $\alpha_3^g$. The latter will be set to zero for the analysis. The matrices $U_g$ then read
\begin{align}\label{eq-UDelta}
    U_g =
    \begin{pmatrix}
    e^{i\alpha_1^g}&0&0\\
    0&e^{i\alpha_2^g}&0\\
    0&0&e^{i\alpha_3^g}
    \end{pmatrix}
    \begin{pmatrix}
    1 & 0 & 0 \\
    0 & c_{23}^g & s_{23}^g \\
    0 & -s_{23}^g & c_{23}^g
    \end{pmatrix}
    \begin{pmatrix}
    c_{13}^g & 0 & s_{13}^ge^{-i \delta^g}  \\
    0 & 1 & 0 \\
    -s_{13}^ge^{i \delta^g} & 0 & c_{13}^g  
    \end{pmatrix}
    \begin{pmatrix}
    c_{12}^g & s_{12}^g & 0 \\
    -s_{12}^g & c_{12}^g & 0 \\
    0 & 0 & 1
    \end{pmatrix}
    \begin{pmatrix}
    e^{i\beta_1^g}&0&0\\
    0&e^{i\beta_2^g}&0\\
    0&0&1
    \end{pmatrix},
\end{align}
where $s_{ij}^g=\sin(\theta_{ij}^g)$, $c_{ij}^g=\cos(\theta_{ij}^g)$, $i,j=1,2,3$. Then the matrices $Y_1$, $Y_3$ and $Y_6$ are given by the relation 
\begin{align}
    Y_1 = Y_3 = Y_6 = Y_\Delta.
\end{align}
The down-type Yukawa matrix $Y_d$ is implemented as a diagonal matrix with real positive entries
\begin{align}
    Y_d=\text{diag}(y_1^d,y_2^d,y_3^d).
\end{align}
The charged lepton Yukawa matrix is for the two models then given by Eq.~\eqref{eq-GUT scale Yukawa ratio 1} and Eq.~\eqref{eq-GUT scale Yukawa ratio 2}, respectively, and implemented accordingly.
\subsection{Parameters and observables}\label{sec-input parameters}
After having discussed the implementation of the Yukawa matrices, we now give the GUT scale input parameters. In total, there are 36 parameters, some of which can be ignored for some parts of the numerical analysis (cf. Section~\ref{sec-fit to experimental data}). These parameters decompose into the GUT scale $M_{\text{GUT}}$, the unified gauge coupling $g_{\text{GUT}}$, the intermediate-scale masses $m_{\Delta_1},$ $m_{\Delta_3},$ $m_{\Delta_6},$ $m_{\phi_1},$ $m_{\phi_2},$ $
m_{\phi_3},$ $m_{\phi_4},$ $m_{\phi_5},$ $m_{t_1},$ $m_{t_2},$ $m_{\Sigma_1},$ $m_{\Sigma_8}$, the scalar coupling constant $\mu$, the singular values $y_1^u,$ $y_2^u,$ $y_3^u,$ $y_1^d,$ $y_2^d,$ $y_3^d,$ $y_1^\Delta,$ $y_2^\Delta,$ $y_3^\Delta$ as well as the angles $\theta_{12}^u,$ $\theta_{13}^u,$ $\theta_{23}^u,$ $\theta_{12}^\Delta,$ $\theta_{13}^\Delta,$ $\theta_{23}^\Delta$ and phases $\delta^u,$ $\beta_1^u,$  $\beta_2^u,$ $\delta^\Delta,$ $\beta_1^\Delta,$ $\beta_2^\Delta$ of the Yukawa matrices. The following list summarizes all the input parameters as well as their allowed ranges:\footnote{The upper bound for the parameters $y_i^\Delta$, where $i=1,2,3$, is discussed in Appendix~\ref{sec-perturbativity of Ydelta above the GUT scale}.}\footnote{The cubic coupling $\mu$ cannot be arbitrarily large. Large values typically drives the potential to a deeper charge-breaking (or color-breaking, depending on the theory) minimum \cite{Frere:1983ag,Alvarez-Gaume:1983drc,Gunion:1987qv,Barroso:2005hc}. This is why we adopt a conservative bound of  $\mu<3m_{\Delta_1}$ for our numerical study.}
\begin{align}\label{eq-input parameters}
M_\text{GUT}&< M_{Pl},  \nonumber\\ m_{\Delta_1},m_{\Delta_3},m_{\Delta_6},m_{\phi_1},m_{\phi_2},
m_{\phi_3},m_{\phi_4},m_{\phi_5},m_{\Sigma_1},m_{\Sigma_8} &\in [500\,\text{GeV},M_\text{GUT}],  \nonumber\\
m_{t_1},m_{t_2}&\in [10^{11}\,\text{GeV},M_{\text{GUT}}], \nonumber\\
\mu&<3m_{\Delta_1}, \nonumber \\
g_{\text{GUT}},\,y_1^u,\,y_2^u,\,y_3^u,\,y_1^d,\,y_2^d,\,y_3^d&\in[0,1],   \\
y_1^\Delta,\,y_2^\Delta,\,y_3^\Delta &\in[0,1.9],   \nonumber\\
\theta_{12}^u,\,\theta_{13}^u,\,\theta_{23}^u,\,\theta_{12}^\Delta,\,\theta_{13}^\Delta,\,\theta_{23}^\Delta &\in[0,\pi/2],   \nonumber\\
\delta^u,\,\beta_1^u,\,\beta_2^u,\,\delta^\Delta,\,\beta_1^\Delta,\,\beta_2^\Delta &\in[-\pi,\pi). \nonumber
\end{align}
The two models are fitted to the following 22 low energy observables (as well as to the nucleon decay rates for 13 different decay channels, see Table~\ref{tab-nucleon decay channels experimental bounds}):
\begin{align}
&g_1,\,g_2,\,g_3\nonumber\\
&y_u,\,y_c,\,y_t,\,y_d,\,y_s,\,y_b,\,\theta_{12}^{\text{CKM}},\,\theta_{13}^{\text{CKM}},\,\theta_{23}^{\text{CKM}},\,\delta^{\text{CKM}},\,y_e,\,y_\mu,\,y_\tau,\,\hspace{0.6cm}\\
&\Delta m_{21}^2,\,\Delta m_{31}^2,\,\theta_{12}^{\text{PMNS}},\,\theta_{13}^{\text{PMNS}},\,\theta_{23}^{\text{PMNS}},\,\delta^{\text{PMNS}}.\nonumber
\end{align}
The GUT scale $M_\text{GUT}$, the unified gauge coupling $g_\text{GUT}$ as well as the intermediate-scale masses are used to fit the SM gauge couplings $g_1$, $g_2$ and $g_3$. Moreover, since the quasi-Yukawa couplings $Y_1$, $Y_3$ and $Y_6$ effect the RG evolution of the SM Yukawa couplings $Y_e$ and $Y_d$, the masses $m_{\Delta_1}$, $m_{\Delta_3}$ and $m_{\Delta_6}$ are further used to fit the singular values of the charged-lepton Yukawa matrix $y_e$, $y_\mu$ and $y_\tau$. The singular values $y_i^u$ and $y_i^d$ ($i=1,2,3$) are used to fit the up- and down-type Yukawa couplings, while the angles $\theta_{12}^u$, $\theta_{13}^u$, $\theta_{23}^u$, and the phase $\delta^u$ are used to fit the CKM parameters. The singular values $y_i^\Delta$ ($i=1,2,3$) together with the trilinear coupling $\mu$ as well as the angles $\theta_{12}^\Delta$, $\theta_{13}^\Delta$, $\theta_{23}^\Delta$, $\delta^\Delta$ are used to fit the neutrino sector. Furthermore, the phases $\beta_1^\Delta$ and $\beta_2^\Delta$ are related to the Majorana phases in the PMNS matrix. Finally, the phases $\beta_1^u$ and $\beta_2^u$ are the so-called GUT phases\ \cite{Ellis:1979hy, Ellis:2019fwf}. They have an impact on the nucleon decay rates, but do not change the predictions for the SM parameters.

\begin{table}[ht!]
\centering
\begin{tabu}{llll}\tabucline[1.1pt]{-} \noalign{\vskip 2mm}
& decay channel  &  $\tau/\mathcal{B}$ [year] & $\Gamma_{\text{partial}}$ [GeV]  
\\\noalign{\vskip 1mm}\hline\noalign{\vskip 2mm}
Proton: & $p\rightarrow \pi^0\,e^+$  &  $>\,2.4\cdot 10^{34}$  &  $>\,8.7\cdot 10^{-67}$  \\\noalign{\vskip 2mm}
&  $p\rightarrow \pi^0\,\mu^+$  &  $>\,1.6\cdot 10^{34}$  &  $>\,1.3\cdot 10^{-66}$  \\\noalign{\vskip 2mm}
&  $p\rightarrow \eta^0\,e^+$  &  $>\,4.1\cdot 10^{33}$  &  $>\,5.1\cdot 10^{-66}$  \\\noalign{\vskip 2mm}
&  $p\rightarrow \eta^0\,\mu^+$  &  $>\,1.2\cdot 10^{33}$  &  $>\,1.7\cdot 10^{-65}$  \\\noalign{\vskip 2mm}
&  $p\rightarrow K^0\,e^+$  &  $>\,1.1\cdot 10^{33}$  &  $>\,1.9\cdot 10^{-65}$  \\ \noalign{\vskip 2mm}
&  $p\rightarrow K^0\,\mu^+$  &  $>\,1.6\cdot 10^{34}$  &  $>\,1.3\cdot 10^{-66}$  \\\noalign{\vskip 2mm}
&  $p\rightarrow \pi^+\,\overline{\nu}$  &  $>\,2.8\cdot 10^{32}$  &  $>\,7.4\cdot 10^{-65}$  \\\noalign{\vskip 2mm}
&  $p\rightarrow K^+\,\overline{\nu}$  &  $>\,6.6\cdot 10^{33}$  &  $>\,3.2\cdot 10^{-66}$ 
\\\noalign{\vskip 1mm}\hline\noalign{\vskip 2mm}
Neutron:  &  $n\rightarrow \pi^-\,e^+$  &  $>\,2.1\cdot 10^{33}$  &  $>\,1.0\cdot 10^{-65}$  \\\noalign{\vskip 2mm}  
&  $n\rightarrow \pi^-\,\mu^+$  &  $>\,9.9\cdot 10^{32}$  &  $>\,2.1\cdot 10^{-65}$  \\\noalign{\vskip 2mm}
&  $n\rightarrow \pi^0\,\overline{\nu}$  &  $>\,9.9\cdot 10^{32}$  &  $>\,2.1\cdot 10^{-65}$  \\\noalign{\vskip 2mm}
&  $n\rightarrow \eta^0\,\overline{\nu}$  &  $>\,5.6\cdot 10^{32}$  &  $>\,3.7\cdot 10^{-65}$  \\\noalign{\vskip 2mm}
&  $n\rightarrow K^0\,\overline{\nu}$  &  $>\,1.2\cdot 10^{32}$  &  $>\,1.7\cdot 10^{-64}$  \\\noalign{\vskip 1mm}
\tabucline[1.1pt]{-}
\end{tabu}
\caption{Current experimental bounds \cite{Dev:2022jbf} on the decay widths $\Gamma_{\text{partial}}$ (respectively lifetime bound $\tau/\mathcal{B} $, where $\mathcal{B}$ is the branching ratio for the decay channel) for various nucleon decay channels at 90 \% confidence level. The data stems from Figure 5-3 of \cite{Brock:2012ogj} and was updated with the findings from \cite{Mine:2016mxy, Takenaka:2020vqy, Bajc:2016qcc, Miura:2016krn, Abe:2014mwa, 1205.6538}.}\label{tab-nucleon decay channels experimental bounds}
\end{table}

\subsection{Fit to experimental data}\label{sec-fit to experimental data}
As described in Section~\ref{sec-implementation} we implement the input parameters given in Eq.~\eqref{eq-input parameters} at the GUT scale $M_\text{GUT}$ and compute the running to the $Z$ scale $M_Z$. The running of the SM gauge couplings is computed at 2-loop, while the running of the (quasi-) Yukawa matrices as well as the effective neutrino mass operators is accounted for at 1-loop. Then, at $M_Z$ the singular values of the SM Yukawa matrices as well as the CKM and PMNS parameters and the neutrino squared mass differences are determined. Furthermore, as described in \cite{Antusch:2021yqe}, the nucleon decay widths for the 13 different decay channels listed in Table~\ref{tab-nucleon decay channels experimental bounds} are computed, using the Mathematica package \texttt{ProtonDecay}\ \cite{Antusch:2020ztu} for the final step of this computation.

At the low scale $M_Z$ we define the total $\chi^2$-function by the sum of the pulls $\chi_i^2$ for each observable $i$
\begin{align}
\chi^2(\vec{x})=\sum_i \chi_i^2(\vec{x}),
\end{align}
where $\vec{x}$ represents a vector which consists of the input parameters listed in Eq.~\eqref{eq-input parameters}. While we use the exact pull $\chi_i^2$ given by NuFIT~5.1~\cite{Esteban:2020cvm} for the observables $\theta_{23}^\text{PMNS}$ and $\delta^\text{PMNS}$, we compute the pull $\chi_i^2$ for all other observables as
\begin{equation}
\chi_i^2(\vec{x})=\left(\frac{f_i(\vec{x})-y_i}{\sigma_{i}}\right)^2,
\end{equation}
where $f_i(\vec{x})$ is our prediction for the observable $i$, while $y_i$ represents the experimental central value and $\sigma_i$ its corresponding standard deviation.\footnote{Note that we take the experimental central value and standard deviations of the SM gauge and Yukawa couplings as well as CKM observables at $M_Z$ from\ \cite{Antusch:2013jca} while the corresponding data for the PMNS observables and neutrino mass squared differences are taken from\ \cite{Esteban:2020cvm}.} Using a differential evolution algorithm the $\chi^2$-function is minimized delivering a benchmark point. Afterwards, starting from that benchmark point, an MCMC analysis is performed using an adaptive Metropolis-Hastings algorithm\ \cite{Metropolis-Hastings-algorithm}. For both models 16 independent chains with $2.5\times 10^5$ data points are computed with a flat prior probability distribution. Using these data points the posterior densities of various observables are calculated.

To simplify the analysis we do not vary all the input parameters listed in Eq.\ \eqref{eq-input parameters} for the minimization procedure. Keeping the masses $m_{\phi_2},$ $m_{\phi_4},$ $m_{\phi_5},$ $m_{t_1},$ $m_{t_2},$ $m_{\Sigma_1}$ and $m_{\Sigma_8}$ at the GUT scale still leaves enough freedom to fit the gauge couplings. Moreover, the correct neutrino mass squared differences can still be obtained if $y_1^\Delta$ is set to zero, which then implies a strong normal hierarchy of the light neutrino masses. Furthermore, since the CKM parameters are almost constant during the RG evolution, the parameters $\theta_{12}^u$, $\theta_{13}^u$, $\theta_{23}^u$, $\delta^u$ are not varied (and the observables $\theta_{12}^\text{CKM}$, $\theta_{13}^\text{CKM}$, $\theta_{23}^\text{CKM}$, $\delta^\text{CKM}$ are not considered in the $\chi^2$-function). Since the GUT phases $\beta_1^u$ and $\beta_2^u$ do not change the predictions for the SM parameters and usually only have a small effect on the nucleon decay predictions in non-SUSY models (see e.g.~\cite{Antusch:2021yqe}) they can be set to zero. Finally, the Majorana phases are also set to zero since they only have a small effect on the running of the SM parameters. In summary, for the fitting procedure the following 20 input parameters are varied
\begin{align}
M_\text{GUT},g_{\text{GUT}},m_{\Delta_1},m_{\Delta_3},m_{\Delta_6},m_{\phi_1},m_{\phi_3},\mu,  \nonumber\\
y_1^u,\,y_2^u,\,y_3^u,\,y_1^d,\,y_2^d,\,y_3^d,y_2^\Delta,\,y_3^\Delta,\theta_{12}^\Delta,\,\theta_{13}^\Delta,\,\theta_{23}^\Delta,\delta^\Delta,
\end{align}
giving a benchmark point. From this benchmark point the MCMC analysis --- for which, on the other hand, all 36 input parameters are varied --- is started.

\section{Results}\label{sec-results}
In this section we present our numerical results for the minimization of the $\chi^2$-function as well as for the MCMC analysis. We are interested in the predictions for the low-scale charged lepton and down-type quark Yukawa ratios which give a direct hint on the viability of the GUT scale Yukawa ratios. Furthermore, we give the predictions of the highest posterior densities (HPD) for the masses of the added scalar fields as well as the nucleon decay widths. All of these predictions may be used to test our models by upcoming experiments.

\subsection{Benchmark points}\label{sec-benchmark points}
Minimizing the $\chi^2$-function under the constraints presented in Section~\ref{sec-fit to experimental data} gives for both models a benchmark point. The GUT scale input parameters for the benchmark points of both models are listed in Table~\ref{tab-input parameters benchmark points}, while Table~\ref{tab-chi squared benchmark points} shows the respective total $\chi^2$ as well as the dominant pulls $\chi_i^2$. With a total $\chi^2$ of 1.86 model 2 gives a better fit than model 1 ($\chi^2=4.35$). In both models the dominant pull comes from the Yukawa coupling of the strange quark with $\chi_{y_s}^2=1.55\,(2.75)$ in model 2 (1). 

Regarding the GUT scale input parameters, the main difference between the two models are the choices of the masses of the intermediate-scale scalar fields. The reason for this is that --- as discussed in Section~\ref{sec-viable GUT scale Yukawa ratios} --- if $m_{\Delta_1}$ is well below the GUT scale, while $m_{\Delta_6}$ is close to the GUT scale the running of $Y_e$ is ``speeded up'' yielding larger charged lepton and down-type quark GUT scale ratios, while conversely, if $m_{\Delta_6}$ is well below the GUT scale, while $m_{\Delta_1}$ is close to the GUT scale the running of $Y_d$ is ``speeded up'' resulting in smaller charged lepton and down-type quark GUT scale ratios. This fixes the masses $m_{\Delta_1}$, $m_{\Delta_3}$ and $m_{\Delta_6}$ in the two models. The remaining masses have then to be chosen such that the gauge couplings can unify which is the reason why they differ in the two models. 

Moreover, an interesting difference between the two models is the chosen GUT scale value for $\theta_{23}^\Delta$ which is, apart from renormalization group effects, equal to the observable $\theta_{23}^\text{PMNS}$. The current experimental 1-$\sigma$ range of this observable consists of two disjoint intervals. The benchmark point of model 1 predicts $\theta_{23}^\text{PMNS}$ to lie in the lower interval, while the benchmark point of model 2 suggests that $\theta_{23}^\text{PMNS}$ should lie in the upper interval. Therefore, a more precise measurement of $\theta_{23}^\text{PMNS}$ could distinguish between the two models (cf.\ Section\ \ref{sec-MCMC}).

\begin{table}[ht]
\centering
\renewcommand{\arraystretch}{1.1}
\begin{tabu}{|c|ccc|ccc|}
\tabucline[1.1pt]{-}
&&Model 1&&&Model 2&\\
\hline
$g_\text{GUT}\;/\;10^{-1}$ & & 6.12 & & & 6.39 & \\ 
$\log_{10}(M_{\text{GUT}}\;/\;\text{GeV})$ & & 15.7 & & & 15.8 &\\
$\log_{10}(m_{\Delta_1}\;/\;\text{GeV})$ & & 15.1 & & & 9.52 &\\ 
$\log_{10}(m_{\Delta_3}\;/\;\text{GeV})$ & & 10.4 & & & 15.0 &\\ 
$\log_{10}(m_{\Delta_6}\;/\;\text{GeV})$ & & 5.79 & & & 13.1 &\\ 
$\log_{10}(m_{\phi_1}\;/\;\text{GeV})$ & & 9.97 & & & 2.70 &\\ 
$\log_{10}(m_{\phi_3}\;/\;\text{GeV})$ & & 6.72 & & & 9.56 &\\ 
$\log_{10}(\mu\;/\;\text{GeV})$ & & 15.6 & & & 4.34 &\\
$y_1^u\;/\;10^{-6}$ & & 2.69 & & & 2.24 &\\
$y_2^u\;/\;10^{-3}$ & & 1.31 & & & 1.10 &\\
$y_3^u\;/\;10^{-1}$ & & 4.05 & & & 3.35 &\\
$y_1^d\;/\;10^{-6}$ & & 6.03 & & & 5.11 & \\ 
$y_2^d\;/\;10^{-4}$ & & 1.39 & & & 0.96 & \\ 
$y_3^d\;/\;10^{-3}$ & & 7.48 & & & 4.82 & \\ 
$y_2^\Delta\;/\;10^{-1}$ & & 1.92 & & & 1.28 & \\ 
$y_3^\Delta\;/\;10^{-1}$ & & 19.0 & & & 9.46 & \\
$\theta^\Delta_{12}\;/\;10^{-1}$ & & 5.81 & & & 5.82 & \\
$\theta^\Delta_{13}\;/\;10^{-1}$ & & 1.15 & & & 1.33 & \\ 
$\theta^\Delta_{23}\;/\;10^{-1}$ & & 6.00 & & & 8.00 & \\ 
$\delta^\Delta$ & & 3.39 & & & 3.39 &\\
\tabucline[1.1pt]{-}
\end{tabu} 
\caption{The GUT scale input parameters of the benchmark points for both models.}\label{tab-input parameters benchmark points}
\end{table}

\begin{table}[ht]
\centering
\renewcommand{\arraystretch}{1.1}
\begin{tabu}{|c|ccccccc|}
\tabucline[1.1pt]{-}
& $\chi^2$ & $\chi^2_{y_s}$ & $\chi^2_{y_b}$ & $\chi^2_{y_\mu}$ & $\chi^2_{y_\tau}$ & $\chi^2_{\theta_{23}^\text{PMNS}}$ & $\chi^2_{\Gamma({p\rightarrow \pi^0e^+})}$ \\
\hline
model 1 & 4.35 & 2.75 & 0.23 & 0.09 & 0.23 & 0.72 & 0.21 \\
model 2 & 1.86 & 1.55 & 0.09 & 0.05 & 0.09 & 0.00 & 0.03 \\
\tabucline[1.1pt]{-}
\end{tabu} 
\caption{The total $\chi^2$ as well as the dominant pulls $\chi_i^2$ for the benchmark points of both models.}\label{tab-chi squared benchmark points}
\end{table}

\subsection{MCMC analysis}\label{sec-MCMC}
We obtain the highest posterior densities of different quantities using an MCMC analysis varying the input parameters around the benchmark points. For this analysis all the input parameters listed in Eq.~\eqref{eq-input parameters} are varied. 

\begin{enumerate}
\item \textbf{Charged lepton and down-type quark Yukawa ratios}\newline
Figure~\ref{fig-HPD Yukawa} shows the 1-$\sigma$ (dark) and 2-$\sigma$ (light) HPD intervals of the charged lepton and down-type quark Yukawa ratios at the low scale $M_Z$ for model 1 (red) and model 2 (green). The experimental 1-$\sigma$ range is indicated by the white region, while the dashed line represents the current experimental central value. As it could already be seen from the benchmark points (cf.\ Section~\ref{sec-benchmark points}), the ratio $\frac{y_\mu}{y_s}$ is predicted to be above the experimental central value in both models. However, there is an intersection between the 1-$\sigma$ HPD interval and the experimentally allowed 1-$\sigma$ region. Future experimental measurements of the three ratios can further test the two models.

\begin{figure}[th]
    \centering
    \includegraphics[width=4.5cm]{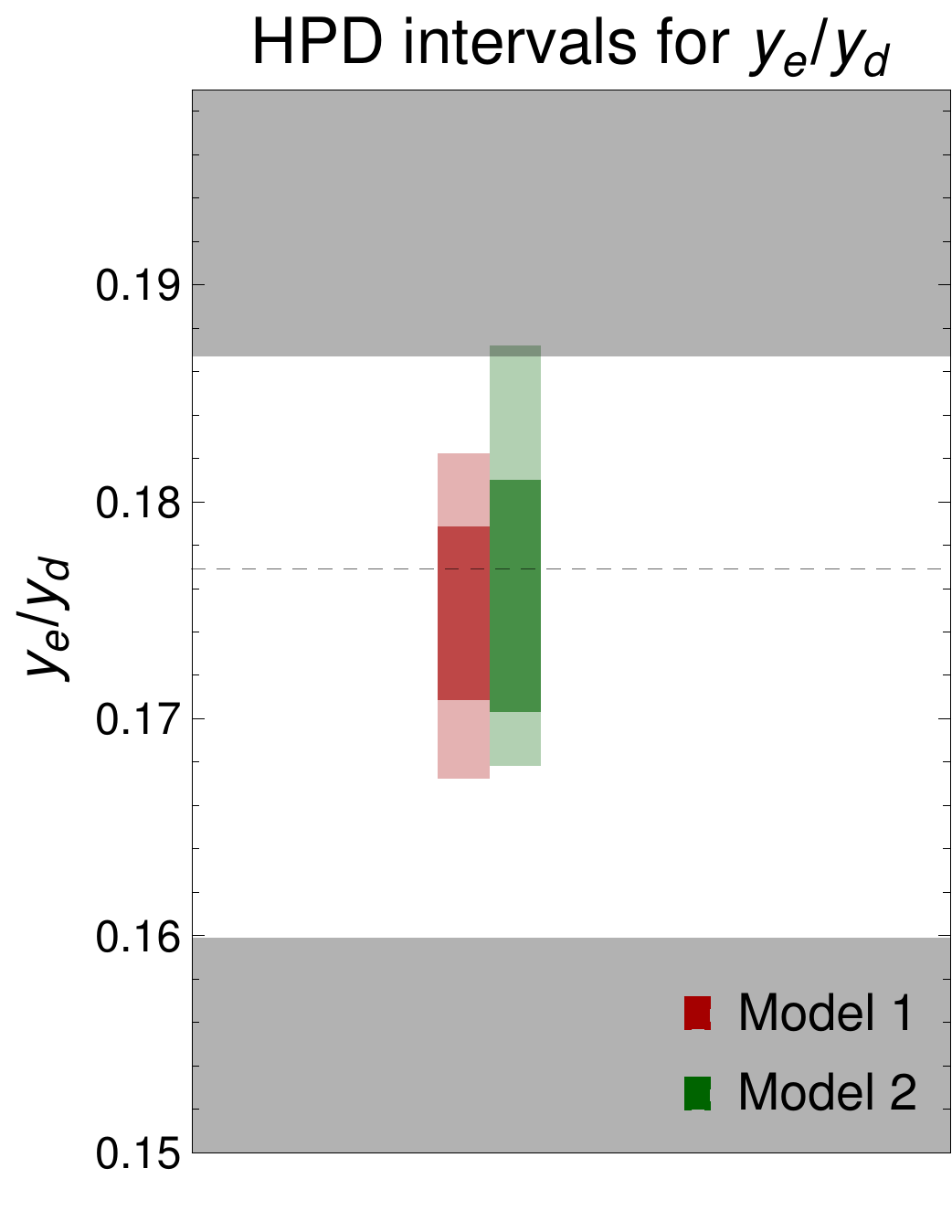}
    \includegraphics[width=4.5cm]{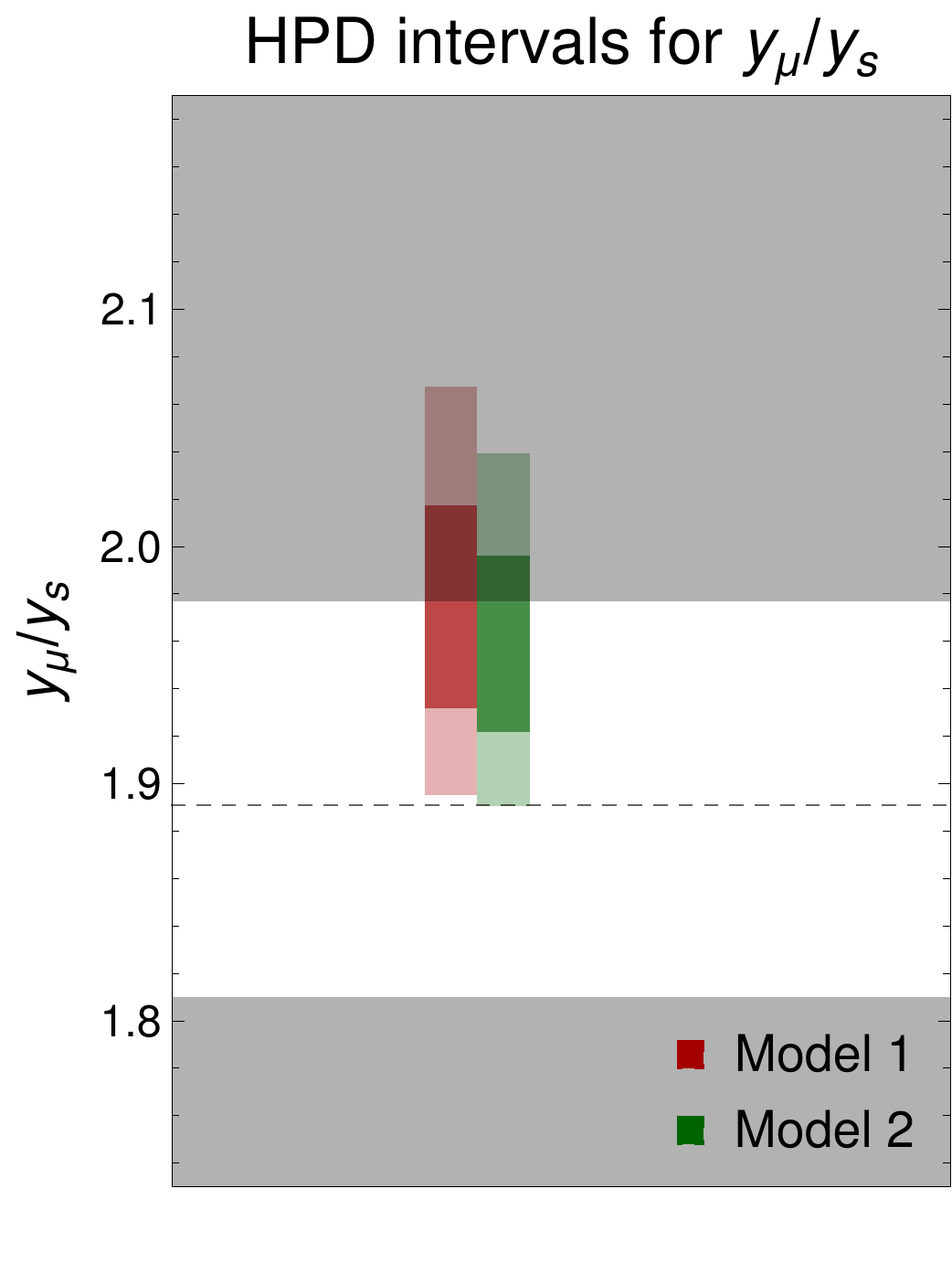}
    \includegraphics[width=4.5cm]{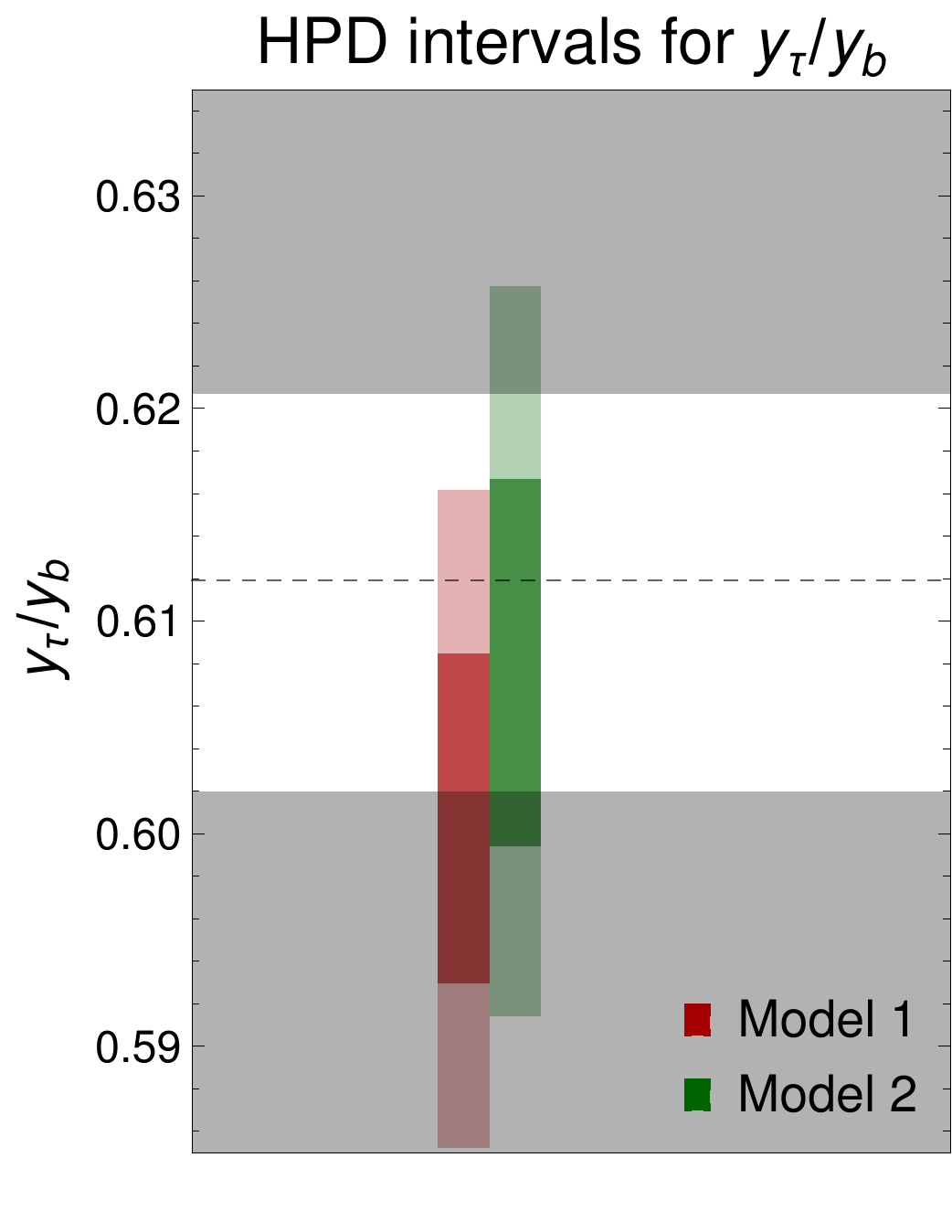}
    \caption{The predicted 1-$\sigma$ and 2-$\sigma$ HPD intervals of the charged lepton and down-type quark Yukawa ratios at $M_Z.$}
    \label{fig-HPD Yukawa}
\end{figure}

\item \textbf{PMNS mixing angle $\theta_{23}^\text{PMNS}$}\newline
As it was pointed out in Section\ \ref{sec-benchmark points} the benchmark points indicate distinct predictions for the observable $\theta_{23}^\text{PMNS}$. Figure\ \ref{fig-HPD theta23} shows that the predicted HPD intervals for model 1 lie within the lower experimentally allowed region, while the predicted HPD intervals of model 2 span over both allowed regions. Therefore, a measurement of $\theta_{23}^\text{PMNS}$ in the upper interval would favor model 2 over model 1.

\begin{figure}
    \centering
    \includegraphics[width=4.5cm]{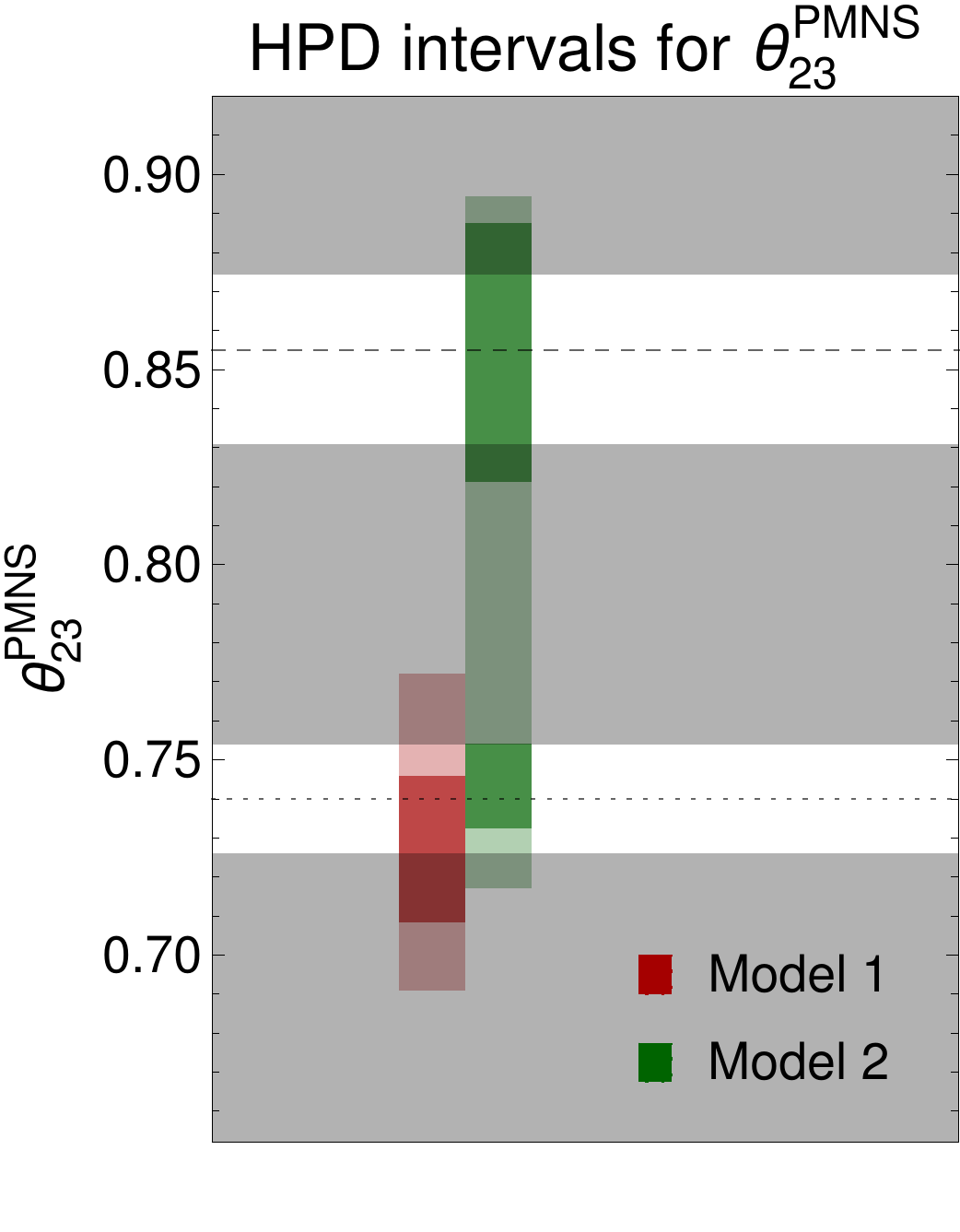}
    \caption{The predicted 1-$\sigma$ and 2-$\sigma$ HPD intervals at $M_Z$ of the PMNS mixing angle $\theta_{23}^\text{PMNS}$.}
    \label{fig-HPD theta23}
\end{figure}

\item \textbf{Intermediate-scale scalar masses}\newline
The prediction for the 1-$\sigma$ and 2-$\sigma$ HPD intervals of the masses of the added scalar fields are presented in Figure~\ref{fig-HPD masses}. The ranges for most of the masses are quite large. However, some of the fields could be observed by future experiments. Especially interesting is the predicted range of $m_{\phi_1}$ in model 2. The upper bound of the predicted 1-$\sigma$ and 2-$\sigma$ HPD interval is given by 13.5\ TeV and 207\ TeV, respectively. Moreover, a possible observation of one of the fields $\Delta_6$, $\phi_1$ or $\phi_3$ could distinguish between the two models.

\item \textbf{Nucleon decay widths}\newline
Figure~\ref{fig-HPD nucleon decay rates} visualizes the predicted 1-$\sigma$ and 2-$\sigma$ HPD intervals for the nucleon decay widths of various decay channels (cf.\ Table\ \ref{tab-nucleon decay channels experimental bounds}). The horizontal blue lines indicate the current experimental bound at a 90\ \% confidence level. For the dominant decay channel,  $p\rightarrow \pi^0 e^+$, also the expected bounds of the planned DUNE\ \cite{Acciarri:2015uup} and Hyper-Kamiokande experiment\ \cite{Abe:2018uyc} are shown. Of particular interest will be the planned Hyper-Kamiokande experiment which will test the full 1-$\sigma$ region of this decay channel predicted by model 1. 

Moreover, as Figure\ \ref{fig-HPD MGUT} shows, model 1 predicts a lower GUT scale than model 2. 
Therefore, model 1 generally predicts larger decay widths than model 2. The 1-$\sigma$ HPD intervals do not intersect for any of the decay channels. This fact can be used to distinguish between the two models if nucleon decay is observed in at least one of the 13 different decay channels. However, as it was pointed out in\ \cite{Antusch:2021yqe} the Yukawa texture has a non-trivial effect on the nucleon decay predictions. This means  that if different Yukawa textures were considered (which still predict the GUT scale relations $\frac{y_\tau}{y_b}=\frac{3}{2}$ and $\frac{y_\mu}{y_s}=\frac{9}{2}$, respectively $\frac{y_\tau}{y_b}=2$ and $\frac{y_\mu}{y_s}=6$) the two scenarios might no-longer be distinguished so easily.

\begin{figure}
    \centering
    \includegraphics[width=4.5cm]{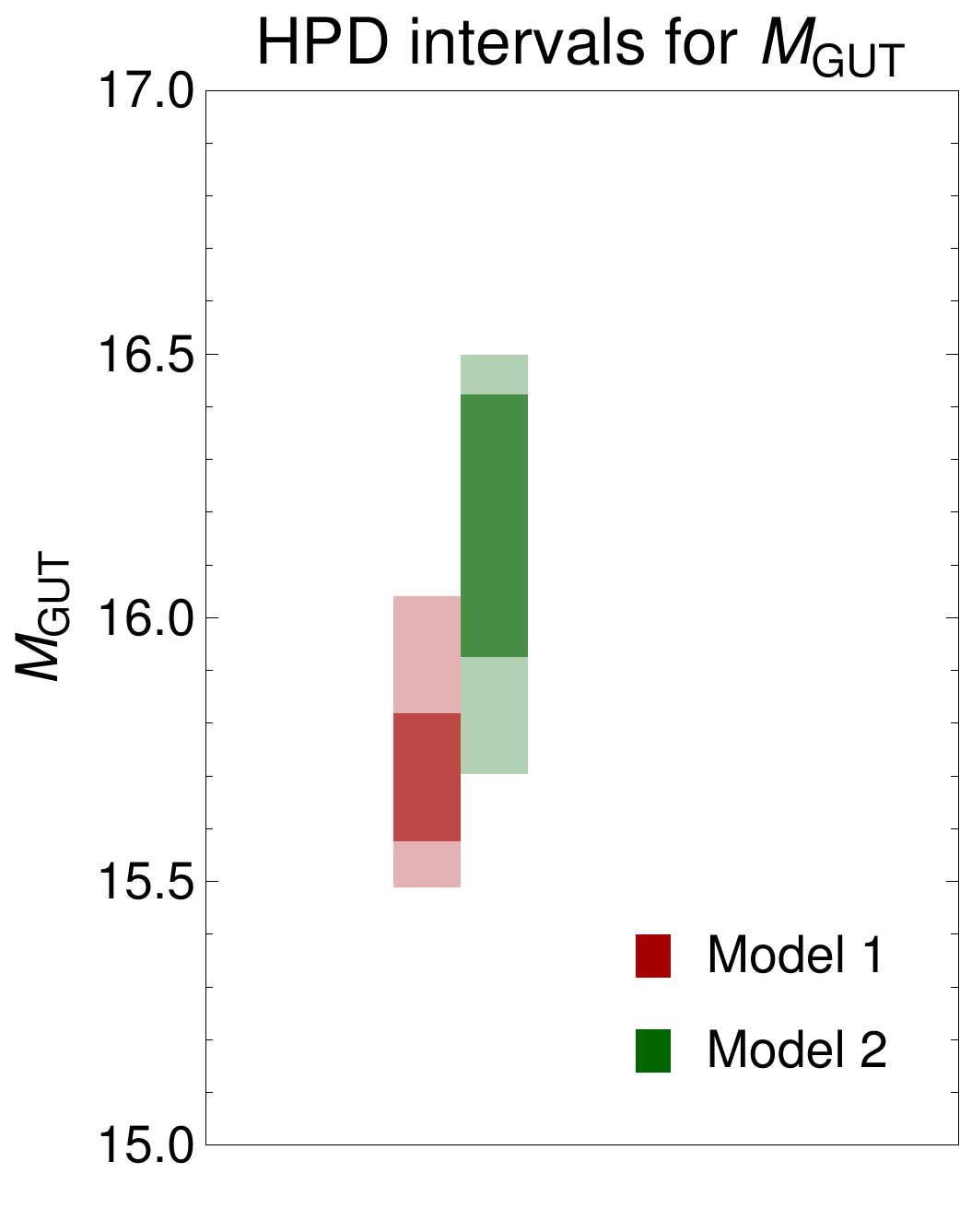}
    \caption{Predicted 1-$\sigma$ and 2-$\sigma$ HPD intervals of the GUT scale $M_\text{GUT}$ for both models.}
    \label{fig-HPD MGUT}
\end{figure}

\end{enumerate}

\begin{figure}[p]
    \centering
    \includegraphics[width=15cm]{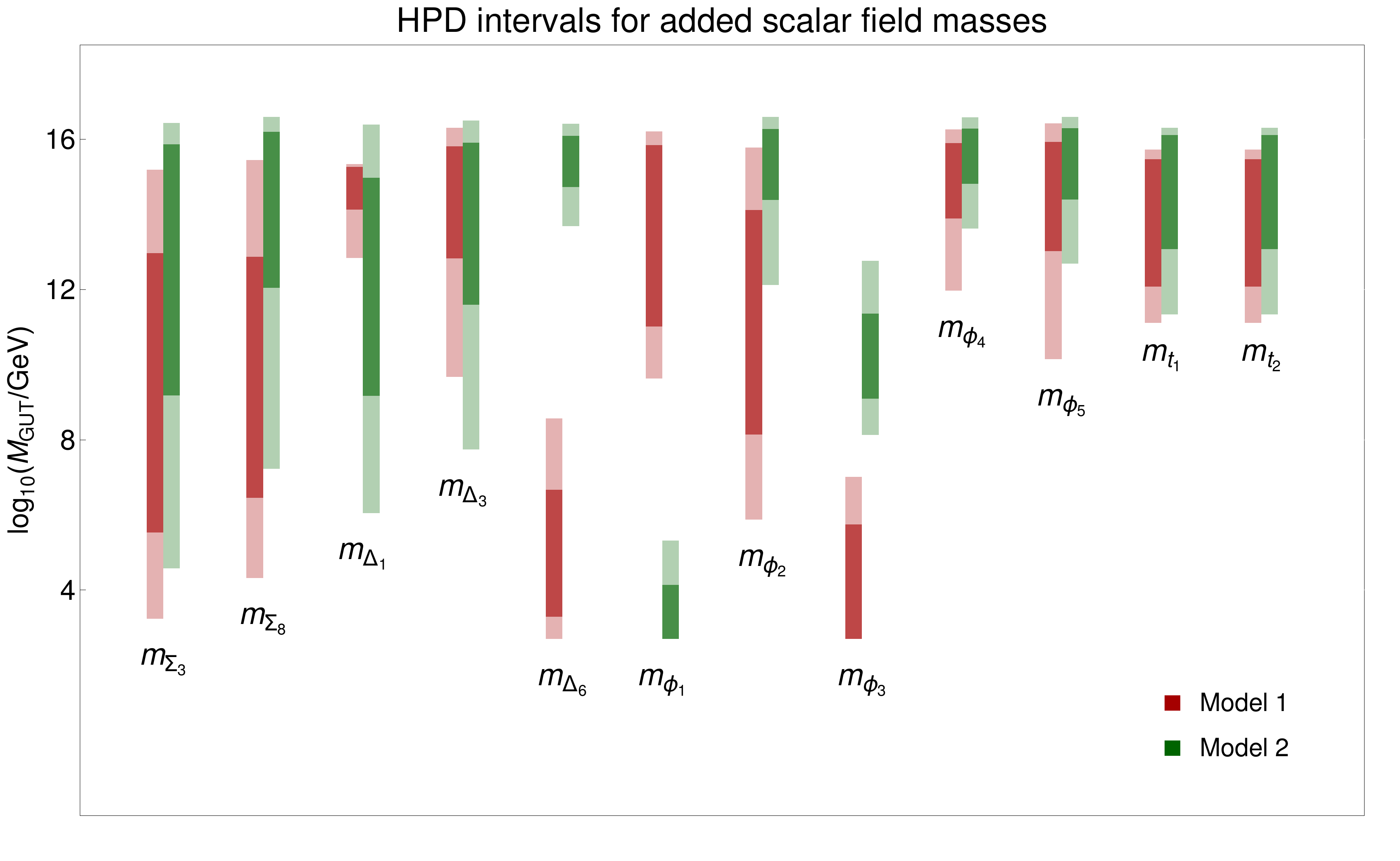}
    \caption{The predicted 1-$\sigma$ and 2-$\sigma$ HPD intervals of the intermediate-scale scalar masses.}
    \label{fig-HPD masses}
\vspace{1cm}
    \includegraphics[width=15cm]{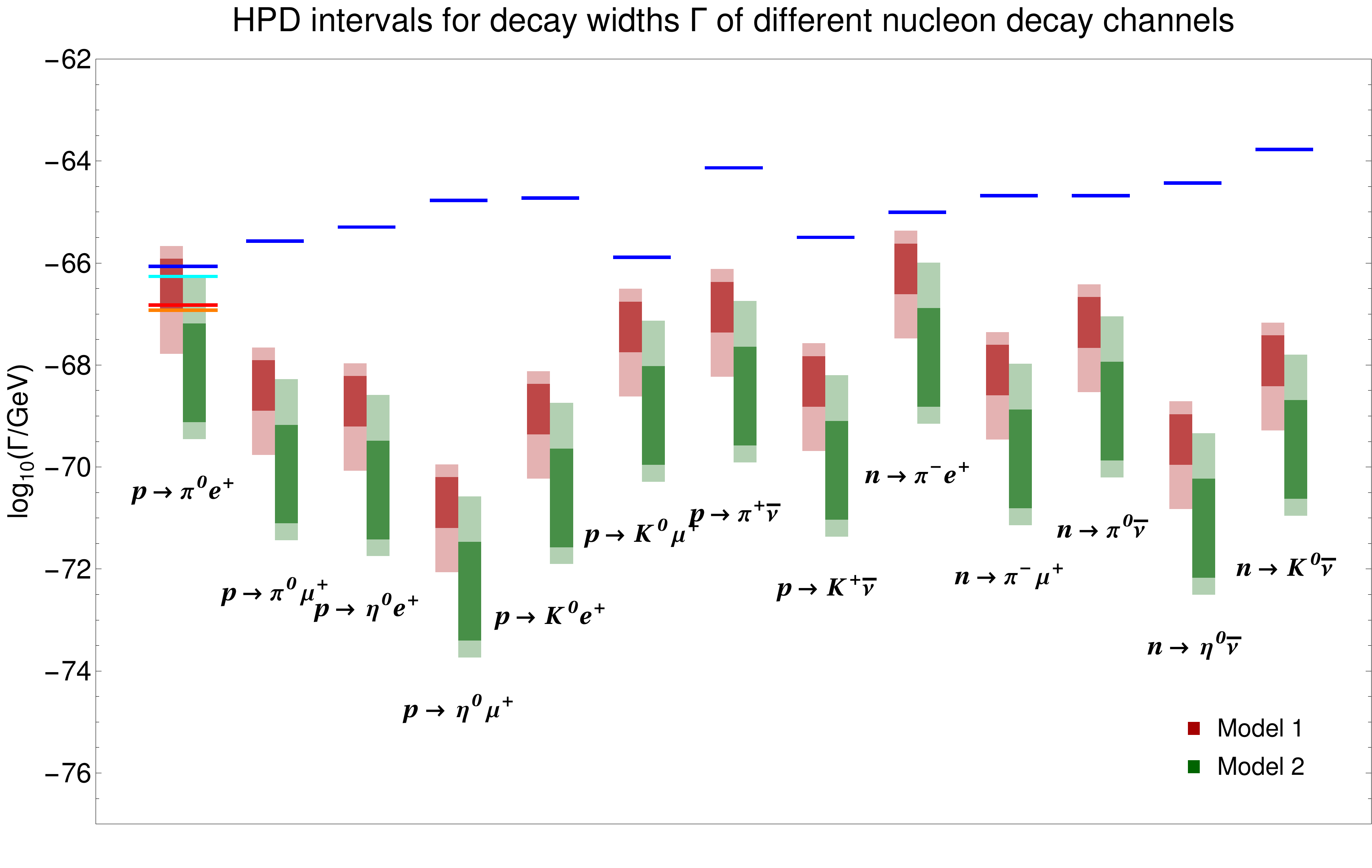}
    \caption{The predicted 1-$\sigma$ and 2-$\sigma$ HPD intervals of the decay widths for various different nucleon decay channels. The current experimental bounds at 90\ \% confident level are indicated by blue lines (cf. Table\ \ref{tab-nucleon decay channels experimental bounds}). Moreover, the future sensitivity of DUNE\ \cite{Acciarri:2015uup} for the decay channel $p\rightarrow \pi^0e^+$ is represented by a cyan line, while the future sensitivity of Hyper-Kamiokande\ \cite{Abe:2018uyc} with one tank (two tanks) for this decay channel is indicated by a red (orange) line.}
    \label{fig-HPD nucleon decay rates}
\end{figure}

It is interesting to note that part of the parameter space, as depicted in Fig.~\ref{fig-HPD masses}, is already ruled out by existing collider searches.  In the following, we briefly summarize the current collider limits. 

The color-octet scalar $\phi_1$ has both neutral $\phi_1^0$ and charged $\phi_1^+$ components. The charged partner provides a stronger bound since it cannot decay to gluons. These states can be efficiently produced via gluon-fusion $pp\to \phi_1^+\phi_1^-, \overline t b \phi_1^+$ at the LHC, which lead to $pp\to \overline t b t \overline b$. The analysis performed in Ref.~\cite{Miralles:2019uzg} shows that the current LHC data provides a lower bound of $m_{\phi_1^+}> 800$ GeV. The neutral components can also be pair-produced through gluon-fusion,  and for our scenario, this would lead to $pp\to \overline b b\overline b b$. LHC currently provides a lower bound~\cite{Hayreter:2017wra} on its mass of $m_{\phi_1^0}>530$ GeV, which is weaker than its charged partner.

Concerning the scalar leptoquark $\phi_3$, which couples predominantly to the second (third) generation quark-lepton pairs for model-1 (model-2), current LHC data limits its mass to be greater than $\gtrsim 1.7$ TeV~\cite{ATLAS:2020dsk} ($\gtrsim 1.4$ TeV~\cite{ATLAS:2019qpq,ATLAS:2021oiz}) from pair-produced leptoquarks decaying into $jj\mu\overline \mu$ (or $bb\tau\overline \tau$) final state.

For both models, the sextet diquark $\Delta_6$ that couples to right-handed down-type quarks $d_{Ri}d_{Rj}$,  has Yukawa coupling of the form $\sim U_\textrm{PMNS}\; \textrm{diag}\{0,\mathcal{O}(0.1),\mathcal{O}(1)\}\; U^T_\textrm{PMNS}$. Subsequently, the coupling to $d_Rd_R$ is somewhat small; however, due to $\mathcal{O}(0.1)$ couplings with $d_Rs_R$ and $s_Rs_R$, it can still be produced through the annihilation  of a pair of quarks via an $s$-channel resonance. By considering  equal diquark couplings of
electromagnetic strength $y_{12,13,23}\sim 0.3$, and using the CMS measurements of narrow dijet resonances at $\sqrt{13}$ TeV~\cite{CMS:2018mgb}, a lower
bound on its mass $\gtrsim 2.5$ TeV is obtained in Ref.~\cite{Pascual-Dias:2020hxo}. This bound is expected to get relaxed in our scenario due to additional non-zero diagonal couplings and the somewhat smaller Yukawa couplings involved in the production. On the other hand, a search for pair-produced resonances in four-jet final states finds diquark mass $\gtrsim 610$ GeV if the decay is into a $b$-quark and a light-quark~\cite{ATLAS:2017jnp}. To obtain the exact bound, however, a dedicated collider study would be required.

Future LHC upgrades have the potential to discover some of these colored particles that could be used to distinguish between these two models.

\section{Conclusions}\label{sec-conclusions}
In this work we considered a rather minimal $SU(5)$ GUT scenario which contains a 45-dimensional as well as a 15-dimensional Higgs representation on top of the Georgi-Glashow $SU(5)$ particle content and in which neutrino masses are generated by a type II seesaw mechanism. We showed that within the concept of \textit{single operator dominance} in the third and second family this scenario can lead to two different GUT scale predictions for charged lepton and down-type Yukawa ratios, namely (i) $\frac{y_\tau}{y_b}=\frac{3}{2}$ and $\frac{y_\mu}{y_s}=\frac{9}{2}$, respectively (ii) $\frac{y_\tau}{y_b}=2$ and $\frac{y_\mu}{y_s}=6$. We investigated the compatibility of both of these two combinations of GUT scale Yukawa ratios with the low-energy experimental data by computing the RG evolution and taking into account the contribution of new arising Yukawa interactions on the beta functions of the charged lepton and down-type quark Yukawa matrices (cf.\ Figure\ \ref{fig-yukawa ratio running}). The outcome of this analysis motivated the formulation of two ``toy models'' which also contain the first family. We investigated various predictions of these ``toy models'' in Section\ \ref{sec-results} and showed that there are several possibilities to distinguish between them in future experiments, namely the nucleon decay rates, the masses of the fields $\Delta_6$, $\phi_1$ or $\phi_3$, which could be discovered at colliders, as well as the prediction for the PMNS 2-3 mixing angle $\theta_{23}^\text{PMNS}$.

\appendix
\appendixpage 
\section{Renormalization group equations of gauge and Yukawa couplings}\label{app-rge gauge and yukawa}
In the following we list the RGEs of the gauge and Yukawa couplings which we have obtained using the Mathematica package \texttt{SARAH} \cite{Staub:2008uz,Staub:2013tta}.

\subsection{Gauge part}

The 2-loop RGEs for gauge couplings ($i,k= 1,2,3$) read
\begin{align}
&\mu\frac{dg_i}{d\mu}=g^3_i\bigg\{ \frac{1}{16\pi^2} a_i+\frac{1}{(16\pi^2)^2}\left( \sum_k  b_{ik}g^2_k  +\beta^{Y}_i  \right)\bigg\},
\end{align}
where $a_i$ ($b_{ik}$) the 1-loop (2-loop) gauge coefficients and $\beta^{Y}_i$ denotes the 2-loop Yukawa contribution. They are given by
\begin{align}
&a_i=a^{SM}_i+\sum_I a^{I}_i\mathcal{H}(\mu,m_{I})\;,\qquad b_{ik}=b_{ik}^{SM}+\sum_I b_{ik}^{I}\mathcal{H}(\mu,m_{I}) \;,
\\
&\beta_i^{Y}=\beta_i^{Y,SM}+\sum_I\beta_i^{Y,I}\mathcal{H}(\mu,m_{I}) \;, \nonumber
\end{align}
with $I$ denoting the respective particle and $\mathcal{H}$ the Heaviside step function which was defined in Eq.\ \eqref{eq-Heaviside step function}. The 1-loop coefficients read
\begin{align}
&a^{SM}_i=  \left(\frac{41}{10}, -\frac{19}{6}, -7\right),\;\;
a^{\Delta_1}_i=  \left(\frac{3}{5}, \frac{2}{3}, 0\right),\;\; 
a^{\Delta_3}_i=  \left(\frac{1}{30}, \frac{1}{2}, \frac{1}{3}\right), \nonumber\\
&a^{\Delta_6}_i=  \left(\frac{8}{15}, 0, \frac{5}{6}\right),\;\;  
a^{\phi_1}_i=  \left(\frac{4}{5}, \frac{4}{3}, 2\right),\;\; 
a^{\phi_2}_i=  \left(\frac{2}{15}, 0, \frac{5}{6}\right), \\
&a^{\phi_3}_i=  \left(\frac{1}{5}, 2, \frac{1}{2}\right),\;\; 
a^{\phi_4}_i=  \left(\frac{49}{30}, \frac{1}{2}, \frac{1}{3}\right),\;\; 
a^{\phi_5}_i=  \left(\frac{1}{15}, 0, \frac{1}{6}\right) ,\nonumber\\
&a^{\phi_6}_i=  \left(\frac{16}{15}, 0, \frac{1}{6}\right),\;\;
a^{\Sigma_3}_i=  \left(0,\frac{1}{3},0\right),\;\; 
a^{\Sigma_8}_i=  \left(0,0,\frac{1}{2}\right), \nonumber
\end{align}
while the 2-loop coefficients are given by
\begin{align}
&b^{SM}_{ik}=\begin{pmatrix}
\frac{199}{50}&\frac{27}{10}&\frac{44}{5}\\
\frac{9}{10}&\frac{35}{6}&12\\
\frac{11}{10}&\frac{9}{2}&-26
\end{pmatrix},\;\;
b^{\Delta_1}_{ik}=\begin{pmatrix}
\frac{108}{25}&\frac{72}{5}&0\\
\frac{24}{5}&\frac{56}{3}&0\\
0&0&0
\end{pmatrix},\;\;
b^{\Delta_3}_{ik}=\begin{pmatrix}
\frac{1}{150}&\frac{3}{10}&\frac{8}{15}\\
\frac{1}{10}&\frac{13}{2}&8\\
\frac{1}{15}&3&\frac{22}{3}
\end{pmatrix}, \nonumber\\
&b^{\Delta_6}_{ik}=\begin{pmatrix}
\frac{128}{75}&0&\frac{64}{3}\\
0&0&0\\
\frac{8}{3}&0&\frac{115}{3}
\end{pmatrix},\;\;
b_{ik}^{\phi_1}=\begin{pmatrix}
\frac{36}{25} & \frac{36}{5} & \frac{144}{5} \\ 
\frac{12}{5} & \frac{52}{3} & 48 \\
\frac{18}{5} & 18 & 84
\end{pmatrix},\;\;
b_{ik}^{\phi_2}=\begin{pmatrix}
\frac{8}{75} & 0 & \frac{16}{3} \\
0&0&0 \\
\frac{2}{3} & 0 & \frac{115}{3}
\end{pmatrix},
\\
&b_{ik}^{\phi_3}=\begin{pmatrix}
\frac{4}{25} & \frac{24}{5} & \frac{16}{5} \\
\frac{8}{5} & 56 & 32 \\
\frac{2}{5} & 12 & 11
\end{pmatrix},\;\;
b_{ik}^{\phi_4}=\begin{pmatrix}
\frac{2401}{150} & \frac{147}{10} & \frac{392}{15} \\ 
\frac{49}{10} & \frac{13}{2} & 8 \\
\frac{49}{15} & 3 & \frac{22}{3}
\end{pmatrix},\;\;
b_{ik}^{\phi_5}=\begin{pmatrix}
\frac{4}{75} & 0 & \frac{16}{15} \\
0 & 0 & 0 \\
\frac{2}{15} & 0 & \frac{11}{3}
\end{pmatrix},\nonumber\\
&b_{ik}^{\phi_6}=\begin{pmatrix}
\frac{1024}{75} & 0 & \frac{256}{15} \\ 
0 & 0 & 0 \\
\frac{32}{15} & 0 & \frac{11}{3}
\end{pmatrix},\;\;
b_{ik}^{\Sigma_3}=\begin{pmatrix}
0&0&0\\
0&\frac{28}{3}&0\\
0&0&0
\end{pmatrix},\;\;
b_{ik}^{\Sigma_8}=\begin{pmatrix}
0&0&0\\
0&0&0\\
0&0&21
\end{pmatrix}.\nonumber
\end{align}
For the 2-loop Yukawa contributions we obtain
\begin{align}
&\beta^{SM,Y}_1=-\frac{17}{10}Tr\left[Y_uY_u^\dagger\right] -\frac{1}{2}Tr\left[Y_dY_d^\dagger\right] -\frac{3}{2}Tr\left[Y_eY_e^\dagger\right], \nonumber
\\
&\beta^{SM,Y}_2=-\frac{3}{2}Tr\left[Y_uY_u^\dagger\right] -\frac{3}{2}Tr\left[Y_dY_d^\dagger\right] -\frac{1}{2}Tr\left[Y_eY_e^\dagger\right],
\\
&\beta^{SM,Y}_3=-2Tr\left[Y_uY_u^\dagger\right] -2Tr\left[Y_dY_d^\dagger\right], \nonumber
\end{align}
and 
\begin{align}
&\bigg\{\beta_1^{Y,\Delta_1},\beta_1^{Y,\Delta_3},\beta_1^{Y,\Delta_6}\bigg\}=\bigg\{-\frac{9}{10}Tr\left[Y_1Y^\ast_1\right], -\frac{13}{10}Tr\left[Y_3Y^\dagger_3\right], -\frac{4}{5}Tr\left[Y_6 Y^\ast_6\right]   \bigg\},\nonumber
\\
&\bigg\{\beta_2^{Y,\Delta_1},\beta_2^{Y,\Delta_3},\beta_2^{Y,\Delta_6}\bigg\}=\bigg\{-\frac{3}{2}Tr\left[Y_1Y^\ast_1\right], -\frac{3}{2}Tr\left[Y_3Y^\dagger_3\right], 0   \bigg\},
\\
&\bigg\{\beta_3^{Y,\Delta_1},\beta_3^{Y,\Delta_3},\beta_3^{Y,\Delta_6}\bigg\}=\bigg\{0, -Tr\left[Y_3Y^\dagger_3\right], -2Tr\left[Y_6 Y^\ast_6\right]   \bigg\}.\nonumber
\end{align}

\subsection{Yukawa part}
The RGEs for Yukawa couplings ($a= u,d,e$) at 1-loop are of the form
\begin{align}
&\mu\frac{dY_a}{d\mu}= \frac{1}{16\pi^2}\bigg\{ \beta^{SM}_a+\sum_I \beta^{I}_a\mathcal{H}(\mu,m_I)\bigg\}  , 
\end{align}
where $\beta_a^{SM}$ are the SM contributions and $\beta_a^I$ are the contributions coming from the new quasi-like Yukawa couplings $Y_1$, $Y_3$ and $Y_6$. The well-known SM contributions read\ \cite{Cheng:1973nv,Machacek:1983fi}
\begin{align}
&\beta^{SM}_u=Y_u\left( -\frac{17}{20}g^2_1-\frac{9}{4}g^2_2-8g^2_3 +T \right)+Y_u \left( \frac{3}{2}Y^\dagger_uY_u-\frac{3}{2}Y^\dagger_dY_d \right)  , \nonumber 
\\
&\beta^{SM}_d=Y_d\left( -\frac{1}{4}g^2_1-\frac{9}{4}g^2_2-8g^2_3 +T \right)+Y_d \left( -\frac{3}{2}Y^\dagger_uY_u+\frac{3}{2}Y^\dagger_dY_d \right)   ,
\\
&\beta^{SM}_e=Y_e\left( -\frac{9}{4}g^2_1-\frac{9}{4}g^2_2+T \right)+Y_e \left( \frac{3}{2}Y^\dagger_eY_e\right), \nonumber
\end{align}
with $T=Tr\left[  3Y^\dagger_uY_u+3Y^\dagger_dY_d+Y^\dagger_eY_e  \right]$. The contributions coming from the new couplings $Y_1$, $Y_3$ and $Y_6$ are
\begin{align}
&\beta^{\Delta_1}_a=\bigg\{ 0,\;0,\; Y_e \left( \frac{3}{2}Y^\ast_1Y_1  \right)\bigg\},\nonumber
\\
&\beta^{\Delta_3}_a=\bigg\{ 0,\;\left(Y_3Y^\dagger_3\right)Y_d,\; Y_e \left( \frac{3}{2}Y^\dagger_3Y_3  \right)\bigg\},
\\
&\beta^{\Delta_6}_a=\bigg\{ 0,\;\left(2Y_6Y^\ast_6 \right)Y_d,\; 0\bigg\}.\nonumber
\end{align}
Furthermore the RGEs of the quasi-like Yukawa couplings read (for $\alpha,\gamma =1, 3, 6$)
\begin{align}
&\mu\frac{dY_\alpha}{d\mu}=\frac{1}{16\pi^2}\bigg\{\beta_{\Delta_\alpha}+\sum_{\gamma\neq \alpha} \delta\beta^{\Delta_\gamma}_{\Delta_\alpha} \mathcal{H}(\mu,m_{\Delta_\gamma})\bigg\} ,   
\end{align}
where we have defined $\beta_{\Delta_\alpha}$ by
\begin{align}
&\beta_{\Delta_1}=Y_1\left( -\frac{9}{10}g^2_1-\frac{9}{2}g^2_2+Tr Y_1Y^\ast_1 \right) +Y_1\left( \frac{1}{2}Y^\dagger_eY_e+3Y^\ast_1Y_1  \right) +\left(\frac{1}{2}Y^T_eY^\ast_e \right)Y_1\;,\nonumber
\\
&\beta_{\Delta_3}= Y_3\left( -\frac{13}{10}g^2_1-\frac{9}{4}g^2_2-4g^2_3+Tr Y_3Y^\dagger_3 \right) +Y_3\left( \frac{1}{2}Y^\dagger_eY_e+\frac{5}{2} Y^\dagger_3Y_3  \right) +\left( Y_dY^\dagger_d \right)Y_3\;,\nonumber
\\
&\beta_{\Delta_6}= Y_6\left( -\frac{2}{5}g^2_1-8g^2_3+Tr Y_6Y^\ast_6 \right) +Y_6\left( Y^\ast_dY_d^T+4 Y^\ast_6Y_6  \right) +\left( Y_dY^\dagger_d \right)Y_6\;,
\end{align}
and $\beta_{\Delta_\alpha}^{\Delta_\gamma}$ as
\begin{align}
&\delta\beta_{\Delta_1}^{\Delta_3}=\frac{3}{2}Y_1Y^\dagger_3 Y_3+\frac{3}{2}Y^T_3Y^\ast_3Y_1\;,\qquad
\delta\beta_{\Delta_1}^{\Delta_6}=0\;,\qquad
\delta\beta_{\Delta_3}^{\Delta_1}=\frac{3}{2}Y_3 Y^\ast_1Y_1\;,\nonumber\\
&\delta\beta_{\Delta_3}^{\Delta_6}=2 Y_6Y^\ast_6Y_3\;,\qquad
\delta\beta_{\Delta_6}^{\Delta_1}=0\;,\qquad
\delta\beta_{\Delta_6}^{\Delta_3}=Y_3Y^\dagger_3Y_6+Y_6Y^\ast_3Y^T_3\;.
\end{align}

\section{Renormalization group evolution of the neutrino mass operator}\label{app-RGE kappa}
The RGE of the dimension five neutrino mass operator $\kappa$ is given by
\begin{align}
    \mu \frac{d\kappa}{d\mu}=\frac{1}{16\pi^2}\left\lbrace\beta_\kappa^{SM}+\sum_{I=\Delta_3,\Delta_6}\beta_\kappa^I\mathcal{H}(\mu,m_{I})\right\rbrace,
 \end{align}
 where the SM contribution $\beta_\kappa^{SM}$ is \cite{Antusch:2001ck}
 \begin{align}
    \beta_\kappa^{SM}=\kappa\left(-\frac{3}{2}Y_e^\dagger Y_e\right) + \left(-\frac{3}{2} Y_e^TY_e^*\right)\kappa+\left(-\frac{3}{2}g_2^2+3T+\lambda\right)\kappa\;.
\end{align}
The neutrino mass operator $\kappa$ is obtained after the field $\Delta_1$ is integrated out below its mass scale $m_{\Delta_1}$ as described in Section\ \ref{sec-neutrino mass generation}. However, if the fields $\Delta_3$ and $\Delta_6$ have masses below $m_{\Delta_1}$, the quasi-like Yukawa couplings $Y_3$ and $Y_6$ potentially affect the running of $\kappa$. Considering all possible Feynman diagrams we have computed this contribution using the method described in\ \cite{Antusch:2001ck}. We obtain $\beta_\kappa^{\Delta_6}=0$ and
\begin{align}
    &\beta_\kappa^{\Delta_3}=\kappa\left(-\frac{3}{2}Y_3^\dagger Y_3\right)+\left(-\frac{3}{2}Y_3^T Y_3^*\right)\kappa\;.
\end{align}

\section{Perturbativity of $Y_\Delta$ above the GUT scale}\label{sec-perturbativity of Ydelta above the GUT scale}
As described in Section\ \ref{sec-viable GUT scale Yukawa ratios} the contribution of the quasi-like Yukawa matrices $Y_1$, $Y_3$ and $Y_6$ on the RG evolution of the charged lepton and down-type quark Yukawa couplings are needed for the compatibility of the GUT scale ratios $\frac{y_\tau}{y_b}=\frac{3}{2}$ and $\frac{y_\mu}{y_s}=\frac{9}{2}$ (respectively $\frac{y_\tau}{y_b}=2$ and $\frac{y_\mu}{y_s}=6$) with the experimental low-energy data. Moreover, in Section\ \ref{sec-benchmark points} we have shown that the largest singular value of $Y_1$, $Y_3$ and $Y_6$ should be of order 1 to obtain a good fit. However, to be consistent with the assumption of \textit{single operator dominance} our ``toy models'' should be perturbative some range above the GUT scale. In order to estimate up to which scale above the GUT scale our ``toy models'' are still perturbative we approximate the 1-loop RGE of $Y_\Delta$ above the GUT scale only taking into account the largest contribution coming from $Y_\Delta$ itself and investigate the running of its singular values. Again, utilizing the method described in\ \cite{Antusch:2001ck}, we obtain 
\begin{align}
    \mu\frac{dY_\Delta}{d\mu}=\frac{1}{16\pi^2}\left(6Y_\Delta Y_\Delta^* Y_\Delta + Tr(Y_\Delta Y_\Delta^*)Y_\Delta\right).
\end{align}
Following the argumentation of\ \cite{Antusch:2021yqe} we say that our ``toy models'' are perturbative as long as the singular values of $Y_\Delta$ are smaller than $\sqrt{4\pi}$ and we require perturbativity of our ``toy models'' up to at least one order of magnitude above the GUT scale. With these assumptions we find an upper bound for the GUT scale value of $y_3^\Delta$ of 1.9 which we used for the analysis in Sections\ \ref{sec-viable GUT scale Yukawa ratios}\ and\ \ref{sec-results}.

\bibliographystyle{style}
\bibliography{references}
\end{document}